\documentclass[12pt]{article}
\pdfoutput=1
\usepackage{color,amsmath,amssymb,exscale,psfrag,epsfig}
\usepackage{cite,color,url}
\usepackage{pdfpages}
\usepackage{graphicx}
\usepackage{slashed}
\usepackage[utf8]{inputenc}
\usepackage[T1]{fontenc}
\usepackage{xcolor}
\usepackage{accents}
\usepackage{centernot}
\usepackage{enumerate}
\usepackage{subfig}
\usepackage{float}
\usepackage{ulem}
\newcommand{\BR}{\text{BR}}
\newcommand{\met}{\ensuremath{E_{\text{T}}^{\text{miss}}}}
\newcommand{\pt}{\ensuremath{p_{\text{T}}}}
\newcommand\tauhad{\ensuremath{\tau_{\text{had}}}}
\usepackage[colorlinks=true
,urlcolor=blue
,anchorcolor=blue
,citecolor=blue
,filecolor=blue
,linkcolor=blue
,menucolor=blue
,linktocpage=true
,pdfproducer=medialab
]{hyperref}
\input epsf
\usepackage{lineno}

\def\bea{\begin{eqnarray}}
\def\eea{\end{eqnarray}}

\definecolor{nicered}{rgb}{0.7,0.1,0.1}
\definecolor{nicegreen}{rgb}{0.1,0.5,0.1}
\hypersetup{colorlinks,citecolor= nicegreen,linkcolor= nicered}

\catcode`\@=11
\def\lsim{\mathrel{\mathpalette\@versim<}}
\def\gsim{\mathrel{\mathpalette\@versim>}}
\def\@versim#1#2{\vcenter{\offinterlineskip
\ialign{$\m@th#1\hfil##\hfil$\crcr#2\crcr\sim\crcr } }}
\catcode`\@=12
\catcode`@=12
\begin{document}
\thispagestyle{empty}
\begin{flushright}
ICAS 034/18
\end{flushright}
\vspace{0.1in}
\begin{center}
{\Large \bf A composite pNGB leptoquark at the LHC} \\
\vspace{0.2in}
{\bf Ezequiel Alvarez$^{(a)\dagger}$,
Leandro Da Rold$^{(b)\ddag}$,\\[1ex]
Aurelio Juste$^{(c,d)\star}$,
Manuel Szewc$^{(a)\diamond}$,
Tamara Vazquez Schroeder$^{(e)\ast}$
}
\vspace{0.2in} \\
{\sl $^{(a)}$ International Center for Advanced Studies (ICAS), UNSAM, Campus Miguelete\\
25 de Mayo y Francia, (1650) Buenos Aires, Argentina }
\\[1ex]
{\sl $^{(b)}$ Centro At\'omico Bariloche, Instituto Balseiro and CONICET\\
Av.\ Bustillo 9500, 8400, S.\ C.\ de Bariloche, Argentina}
\\[1ex]
{\sl $^{(c)}$ 
Institut de F\'isica d'Altes Energies (IFAE), Edifici Cn, Facultat de Ciencies,\\
Universitat Aut\`onoma de Barcelona, E-08193 Bellaterra, Barcelona, Spain}
\\[1ex]
{\sl $^{(d)}$ 
Instituci\'o Catalana de Recerca i Estudis Avan\c{c}ats (ICREA), E-08010 Barcelona, Spain}
\\[1ex]
{\sl $^{(e)}$ 
CERN, CH-1211 Geneva, Switzerland}
\end{center}
\vspace{0.1in}

\begin{abstract}
The measurements of $R_K^{(*)}$ and $R_{D}^{(*)}$ by BaBar, Belle and the LHCb collaborations could be showing a hint of lepton flavor universality violation that can be accommodated by the presence of suitable leptoquarks at the TeV scale. 
We consider an effective description, with leptoquarks arising as composite pseudo Nambu-Goldstone bosons, as well as anarchic partial compositeness of the SM fermions.
Considering the $R_K^{(*)}$ anomaly within this framework, we study pair production of $S_3\sim(\bar 3,3)_{1/3}$ at the LHC. We focus on the component $S_3^{1/3}$ of the triplet, which decays predominantly into $t\tau$ and $b\nu$, and study the bounds from existing searches at $\sqrt{s}=13$ TeV at the LHC. We find that sbottom searches in the $b\bar{b}+\met$ final state best explore the region in parameter space preferred by our model and currently exclude $S_3^{1/3}$ masses up to $\sim$1 TeV.  
Additional searches, considering the $t\tau$ and $t\mu$ decay modes, are required to probe the full physical parameter space. 
In this paper we also recast existing studies on direct leptoquark searches in the $t\tau t\tau$ channel and SM $t\bar{t}t\bar{t}$ searches, and obtain the regions in parameter space currently excluded.  Practically the whole physical parameter space is currently excluded for masses up to $\sim$0.8 TeV, which could be extended up to $\sim$1 TeV with the full Run 3 dataset. We conclude that pair production searches for this leptoquark can benefit from considering the final state $t \tau b +\met$, where the largest branching ratio is expected.  We appraise that future explorations of leptoquarks explaining the B-anomalies with masses beyond the TeV should also consider single and non-resonant production in order to extend the mass reach.
\end{abstract}

\vspace*{2mm}
\noindent {\footnotesize E-mail:
{\tt 
$\dagger$ \href{mailto:sequi@unsam.edu.ar}{sequi@unsam.edu.ar},
$\ddag$ \href{mailto:daroldl@cab.cnea.gov.ar}{daroldl@cab.cnea.gov.ar},
$\star$ \href{mailto:juste@ifae.es}{juste@ifae.es},\\
$\diamond$ \href{mailto:mszewc@unsam.edu.ar}{mszewc@unsam.edu.ar}
$\ast$ \href{mailto:tamara.vazquez.schroeder@cern.ch}{tamara.vazquez.schroeder@cern.ch},
}}


\section{Introduction}
\label{section:1}
Since the birth of the Standard Model (SM), Lepton Flavor Universality (LFU) has been an outstanding feature of particle physics. It is one of the milestones in Fermi's Universal Theory, which is captured by the gauge interactions of the SM, and only broken within the SM in the Yukawa interactions.
Since the establishment of the SM, there has been much interest in searching for LFU violation.  However, only in recent years it has become possible to explore interactions that involve heavy quarks and heavy leptons, i.e. the two sectors characterized by a large flavor universality violation in their Yukawa interactions, and thus where one could expect to find new sources of flavor universality violation.  

Over the last few years, results from B-factories and the LHCb experiment show deviations of $\sim$2--3$\sigma$ in the ratios $R_K^{(*)}$ and $R_D^{(*)}$ \cite{Aaij:2014ora, Aaij:2015yra, Huschle:2015rga,Aaij:2017vbb}, where accurate tests of LFU can be performed.  This has been a subject of intense study in the literature where, mostly within an Effective Field Theory (EFT) approach, one can identify several possible New Physics (NP) explanations \cite{Buttazzo:2017ixm, Hiller:2017bzc, Calibbi:2015kma, Becirevic:2015asa, Dorsner:2017ufx, Crivellin:2017zlb} for the discrepancies.  
Among them, a leptoquark~\cite{Pati:1973uk} arises as one of the favored explanations~\cite{Becirevic:2016yqi, Hiller:2017bzc, Kosnik:2012dj}.  In particular, considering the $R_K^{(*)}$ anomaly, there are three possible leptoquarks that can accommodate the observed results \cite{Hiller:2017bzc}:
\begin{eqnarray}
S_3 \sim (\bar 3, 3)_{1/3}\ ,\quad 
V_1 \sim (3,1)_{2/3}\ ,\quad
V_3 \sim (3,3)_{-2/3}\ , \nonumber
\end{eqnarray}
where $S$ and $V$ denote spin 0 and 1, respectively, and the numbers indicate the corresponding representation under the SM group $SU(3)_C \otimes SU(2)_L \otimes U(1)_Y$.  Recent studies~\cite{Becirevic:2016yqi,Fajfer:2017lzq} have focused in particular on the $S_3$ leptoquark as a favored explanation of the $R_K^{(*)}$ anomaly.  Although recent works \cite{Angelescu:2018tyl,Buttazzo:2017ixm,Crivellin:2017zlb} point out that the $V_1$ option is favored as a single-leptoquark explanation of both $R_K^{(*)}$ and $R_D^{(*)}$ anomalies, along this article we consider a scalar option to explain the $R_K^{(*)}$ anomaly, for the reasons described below.
The combined effect of two scalar leptoquarks, $S_3$ and $S_1 \sim (\bar 3, 1)_{1/3}$, can also provide an explanation for both anomalies~\cite{Crivellin:2017zlb,Becirevic:2018afm,Marzocca:2018wcf}.

In this work we explore the possibility that the leptoquark responsible for the $R_K^{(*)}$ anomaly is a pseudo Nambu-Goldstone boson (pNGB) of a new sector, and we therefore study the case of $S_3$.  
This choice allows a splitting between the mass of $S_3$ from the masses of other NP states, which can be made naturally heavier and whose effect can be neglected, in particular in electroweak precision tests (EWPT). A suitable framework for this scenario is a new strongly coupled sector with resonances heavier than a few TeV \cite{Gripaios:2009dq, Gripaios:2014tna,Marzocca:2018wcf}, and with a global symmetry spontaneously broken by the strong dynamics, generating a set of pNGB's. Within the pNGB's there is the state $S_3$, as well as a composite Higgs boson. There might exist also additional light scalars, but we assume their phenomenological impact in the context of this study can be neglected. We also assume anarchic partial compositeness, which leads to a well defined pattern of couplings, although some departures will be allowed to cover a broader region of couplings.

In particular, we focus on the component of $S_3$ with electric charge $1/3$, denoted $S_3^{1/3}$. The main decay modes are $S_3^{+1/3} \to \bar t\, \ell^+$ and $\bar b \bar \nu$, with a preference for second and third generation leptons.  We consider several recent LHC searches that are sensitive to the presence of the $S_3^{1/3}$ state and derive constraints in its branching ratios as a function of mass.

This paper is organized as follows. In Section~\ref{section:2} we present an effective description that allows to model the leptoquarks as composite states; an example of a coset that can lead to a Higgs and leptoquarks as composite pNGB's will be available in Ref.~\cite{leandro}, also see for example Ref.~\cite{Gripaios:2014tna} and \cite{Marzocca:2018wcf}. We discuss the phenomenology of the model in Section~\ref{section:3}, where we combine the model predictions with the available B-physics anomalies results.  In Section \ref{section:4} we discuss the re-casting of three LHC searches sensitive to this leptoquark, and the results are presented in Section \ref{section:5}. Finally, in Section \ref{section:6} we provide a summary and outlook.

\section{The model}
\label{section:2}
As shown in Ref.~\cite{Hiller:2017bzc}, the tree-level exchange of scalar or vector leptoquarks at the TeV scale generate dimension-six operators with Wilson coefficients of appropriate size to accommodate the deviations observed in LFU observables in B-meson decays. A very interesting possibility is to consider a new strongly coupled sector that, besides the Higgs boson, generates resonances at the TeV scale, some of which are leptoquarks. 
As we will show below, in the context of anarchic partial compositeness, the effect in $R_{K^{(*)}}$ prefers a leptoquark mass $\lesssim$ TeV. However, EWPT in general require the masses of resonances to be $m_*\gtrsim$~few~TeV, thus a separation between the lightest leptoquark mass and $m_*$ is required. This splitting can be obtained if the leptoquark involved in $R_{K^{*}}$ is a pseudo Nambu-Goldstone boson (pNGB), since in this case its mass is given by $m_{\rm pNGB} \simeq \frac{g_{SM}}{4\pi}m_*$, with $g_{\rm SM}$ a SM coupling, as the gauge couplings or the top quark Yukawa coupling. Therefore, in the following we will consider a scenario where the $S_3$ leptoquark emerges 
as a pNGB of the new strongly interacting sector.

In this section we describe a model that includes leptoquarks and allows studying the LHC phenomenology of $S_3$. We consider a strongly coupled field theory (SCFT) that produces resonances with masses $m_*\gtrsim$~few~TeV. The SCFT has a global symmetry $G$, spontaneously broken by the strong dynamics to a subgroup $H$ containing the SM gauge group. The Higgs boson and the $S_3$ leptoquark are pNGBs associated to this spontaneous breaking of global symmetries.\footnote{Although the Higgs boson will not play any role in the phenomenology that we will study, this assumption determines the size of the couplings between the SM fermions and the SCFT.} There might be other pNGB states in the coset $G/H$, leptoquarks or not, but we will assume that their effect is subdominant the phenomenology that we will study, compared with the effect of $S_3$.\footnote{See Ref.~\cite{Gripaios:2014tna} for a realization with a factorizable minimal group. The case of a simple group will be presented elsewhere.} The flavor structure of the SCFT is taken to be anarchic~\cite{Gherghetta:2000qt,Csaki:2008zd}, meaning that there are no preferred directions in flavor space, and thus all the coefficients of tensors in flavor space are of the same order. 

The SM states are elementary, with the SM gauge fields weakly gauging a subgroup of the global symmetries of the SCFT. The SM fermions have linear interactions with the SCFT:
${\cal L}\supset \omega\ \bar\psi_{\rm SM} {\cal O}_{\rm SCFT}$, 
where $\omega$ is the coupling at the UV scale at which this interaction is defined. Bilinear couplings will be taken to be subdominant compared with the effect of the linear ones, and thus not considered in the following. All these interactions explicitly break the global symmetry of the SCFT, resulting in a potential being induced at loop level for the resulting pNGBs. The top quark dominates the contributions to the potential, and it can trigger the breaking of the electroweak symmetry. We will not discuss the details of the potential in this work.

Rather than attempting to develop a fundamental theory, we will consider instead an effective theory describing the resonances and their interactions. For simplicity we will adopt the one scale and one coupling description of the resonances~\cite{Giudice:2007fh}, defining $g_*$ as the typical size of the coupling between the resonances, that is assumed to be larger than the SM couplings but in the perturbative regime, roughly $1<g_*<5$. At low energies the linear interactions of the SM fermions lead to mixing with the resonances of the SCFT:
\begin{equation}
{\cal L}_{\rm eff}\supset \lambda\ f\ \bar\psi_{\rm SM} \Psi_{\rm SCFT} \ , \label{eq-Lpc}	
\end{equation}
where $\lambda$ is the coupling at the scale $m_*$, $f=m_*/g_*$ is the NGB decay constant, and $\Psi_{SCFT}$ is a fermionic vector-like resonance of the composite sector. This mechanism is usually known as partial compositeness~\cite{Kaplan:1991dc}, since the mass eigenstates are a mixture of elementary and composite states, with degree of compositeness $\epsilon\sim\lambda/g_*$. Fermionic resonances and the Higgs have Yukawa interactions: ${\cal L}_{\rm eff}\supset g_* c_{ij}\bar\Psi_{\rm SCFT}H\Psi'_{\rm SCFT}$, with $c_{ij}$ anarchic coefficients in flavor space of ${\cal O}(1)$.

In the case of an anarchic SCFT, linear interactions lead to what is usually referred to as ``anarchic partial compositeness'' (APC). Assuming that the energy evolution of the linear couplings is driven by the dimension of the SCFT operator, as well as a separation between the UV scale of Eq.~(\ref{eq-Lpc}) and $m_*$, it is possible to obtain a hierarchy of mixing of the SM fermions. After electroweak symmetry breaking, the mixing leads to interactions with the Higgs boson that generate masses for the SM fermions. Their Yukawa couplings have the the structure: $y_{ij}\simeq g_*\epsilon_{Li}\epsilon_{Rj}c_{ij}$, where $i,j$ are generation indices and $\epsilon_{Li,Ri}$ is the degree of compositeness of the Left-handed (LH) or Right-handed (RH) quirality. Large masses, as in the case of the top quark, can be obtained by taking the left-handed (LH) and right-handed (RH) mixing to be of ${\cal O}(1)$, meaning that the top quark has a large degree of compositeness, $\epsilon_{t_{L,R}}\sim{\cal O}(1)$. In contrast, tiny masses can be obtained by taking the mixing of one of the chiralities to be small, as in the case of the light quarks and leptons. A hierarchy of mixing of the LH quarks can also lead to the CKM matrix. The scenario of APC also provides a very economic mechanism to satisfy most of the flavor bounds present in Composite Higgs Models~\cite{Csaki:2008zd,KerenZur:2012fr}.\footnote{The most important constraints arise from the kaon system and electromagnetic dipole moments, although some solutions have been proposed~\cite{Bauer:2011ah,DaRold:2017dbr,DaRold:2017xdm}.}

The masses and weak mixing angles of the SM quarks can be reproduced by taking:
\begin{align}
&\epsilon_{q_1}\sim\lambda_C^3 \epsilon_{q_3}\ , 
\qquad 
\epsilon_{u_3}\sim \frac{y_t^{\rm SM}}{g_*\epsilon_{q_3}} \ ,
\qquad
\epsilon_{u_2}\sim \frac{y_c^{\rm SM}}{\lambda_C^2g_*\epsilon_{q_3}}  \ , 
\qquad
\epsilon_{u_1}\sim \frac{y_u^{\rm SM}}{\lambda_C^3g_*\epsilon_{q_3}} \ , 
\nonumber \\
&\epsilon_{q_2}\sim\lambda_C^2 \epsilon_{q_3}\ ,
\qquad
\epsilon_{d_3}\sim \frac{y_b^{\rm SM}}{g_*\epsilon_{q_3}} \ ,
\qquad
\epsilon_{d_2}\sim \frac{y_s^{\rm SM}}{\lambda_C^2g_*\epsilon_{q_3}} \ ,
\qquad
\epsilon_{d_1}\sim \frac{y_d^{\rm SM}}{\lambda_C^3g_*\epsilon_{q_3}} \ ,
\label{eq-mixqR}
\end{align}
where $\epsilon_{q_i}$ corresponds to the LH doublet of $i$th generation, whereas $\epsilon_{u_i}$ and $\epsilon_{d_i}$ to the RH singlets, $\lambda_C$ is the Cabibbo angle, and $y^{\rm SM}$ are the Yukawa couplings of the SM. The degree of compositeness of all the quarks is determined by physical quantities up to a common factor $1/(g_*\epsilon_{q_3})$. 

The lepton sector depends on the realization of neutrino masses. We will assume that the angles of the PMNS matrix are generated by the matrix diagonalizing the neutrino mass matrix. For the charged leptons we will take hierarchical mixing of the same size for both chiralities of each generation. This choice minimizes the constraints from flavor violating transitions in the lepton sector~\cite{Panico:2015jxa}. Therefore, we assume:
\begin{equation}
\epsilon_{\ell_j}\sim\epsilon_{e_j}\sim\sqrt{\frac{m_{e_j}}{ g_* v}}\ , \qquad j=1,2,3 \ ,
\label{eq-coupling}
\end{equation}
where $v$ stands for the vacuum expectation value (vev) of the Higgs field.

The interactions between the SM fermions and the resonances require insertions of the mixing. Although the SCFT is anarchic, the couplings with the resonances are not, since they are mediated by the hierarchical structure of the mixing. Interactions involving two SM fermions, $\psi$ and $\psi'$, as well as a spin zero or one resonance, are expected to be of order~$\epsilon_\psi g_*\epsilon_\psi'$.
Roughly speaking, heavy SM fermions, mainly the top quark, but also the LH bottom quark, and eventually the tau lepton, will have sizable couplings, whereas the coupling of the light SM fermions will in general be suppressed. 

We are interested in the following interactions of $S_3$~\cite{Dorsner:2016wpm}
\begin{equation}
{\cal L}_{\rm int}\supset y\ \bar{q^c_L} (\tau\cdot S_3) \ell_L \ +\ {\rm h.c.} \ , \label{eq-LS3-1}
\end{equation}
where the coupling $y$ has the structure estimated in the previous paragraph, and generation indices are understood. $\tau$ is a shorthand for the matrices $\tau_j=\sigma_j/2$, contracted with the three components $S_3^j$. By using Eqs.~(\ref{eq-mixqR})--(\ref{eq-coupling}), an estimate of the couplings $y$ is given by
\begin{equation}
y\simeq \epsilon_{q_3}\sqrt{g_*}
\left(\begin{array}{ccc}
\lambda_C^3\sqrt{m_e/v} & \lambda_C^3\sqrt{m_\mu/v} & \lambda_C^3\sqrt{m_\tau/v} \\
\lambda_C^2\sqrt{m_e/v} & \lambda_C^2\sqrt{m_\mu/v} & \lambda_C^2\sqrt{m_\tau/v} \\
\sqrt{m_e/v} & \sqrt{m_\mu/v} & \sqrt{m_\tau/v} 
\end{array}
\right) \ ,
\end{equation}
where each coefficient of the matrix must be multiplied by an independent factor of ${\cal O}(1)$.
A numerical estimate of $y$, up to the factor $1/(\epsilon_{q_3}\sqrt{g_*})$, can be found in Ref.~\cite{Hiller:2017bzc}. 

It is useful to define an electric charge eigenstate basis: $S_3^{+4/3}=(S_3^1-iS_3^2)/\sqrt{2}$, $S_3^{+1/3}=S_3^3$ and $S_3^{-2/3}=(S_3^1+iS_3^2)/\sqrt{2}$. Expanding Eq.~(\ref{eq-LS3-1}) in components and rotating to the mass basis of fermions~\cite{Dorsner:2016wpm}:
\begin{align}
{\cal L}_{\rm int}\supset & -\sqrt{2}\ y_{jk}\ \bar{d^c_L}^j\ S_3^{+4/3}\ e_L^k + \sqrt{2}\ (V^tyU)_{jk}\ \bar{u^c_L}^j\ S_3^{-2/3}\ \nu_L^k \nonumber
\\
&-\ S_3^{+1/3}\ [(yU)_{jk}\ \bar{d^c_L}^j\ \nu_L^k+(V^ty)_{jk}\ \bar{u^c_L}^j\ e_L^k] \ +\ {\rm h.c.} \  ,
\label{eq-LS3-2}
\end{align}
where $V$ and $U$ are the CKM and PMNS matrices.

The SM gauge symmetry is compatible with the presence of $S_3qq$ interactions, which can mediate proton decay. The presence of these interactions with leptoquark masses of order TeV would rule out the present scenario. We assume that there exists an additional symmetry that forbids $S_3qq$ interactions~\cite{Hiller:2017bzc}, while allowing $S_3q\ell$ interactions. This can be achieved with a discrete $Z_2$ symmetry, e.g by assigning an odd parity ($-$) to $S_3$ and $q$, and an even parity ($+$) to $\ell$.

The $S_3$ leptoquark also interacts with the Higgs boson. It is interesting to study these interactions because they can split the masses of the components of $S_3$. Considering up to dimension-four operators, there are two independent terms of type $H^2S_3^2$~\cite{Dorsner:2016wpm}. Evaluating the Higgs field on its vev they induce splitting between $S_3^j$. In the presence of other leptoquarks, as $S_1$, $\tilde S_2\sim(3,2)_{1/6}$ and $\hat S_2\sim(3,2)_{-5/6}$, new effects are present at the level of dimension-four operators. The splitting between components of a given multiplet is of order $v$, thus $\Delta m_S\sim {\cal O}(100\ \mbox{GeV})$ can be expected.~\footnote{We thank F. Lamagna for pointing us the splitting for charged $S_3$, correcting a mistake in the original version of this work.}

Given this description, and the corresponding estimates of the couplings, we have the tools to study the phenomenology of $S_3$ at the LHC.

\section{Phenomenology}
\label{section:3}
In the previous section we have presented an effective description of a theory containing a number of new particles.
Among those, the pNGBs are the ones most accessible experimentally, owing to their lower mass. In this section we review the main features of the phenomenology of the $S_3$ leptoquark, which transforms as $\sim (\bar 3, 3)_{1/3}$.  

If the LFU anomalies in B-physics result from the exchange of a new heavy resonance that is off-shell, then it could manifest itself 
as on-shell production at the LHC, provided it is kinematically accessible. Otherwise, it would be hard to detect this new resonance 
at the ATLAS and CMS experiments, since off-shell effects are more difficult to observe than in B-physics experiments.

The three members of the $S_3$ triplet have different decay modes and thus different phenomenology.  
According to their interactions described in the previous section, we find that the main decay modes include a third-generation quark and are
\begin{eqnarray}
S_3^{-2/3} &\to& \sum_{i} \bar t \bar \nu_i \quad \propto \quad |y_{33}|^2+|y_{32}|^2 \ , \label{decay1} \\
S_3^{+1/3} &\to& \bar t \tau^+, \ \bar t \mu^+, \sum_{i}\bar b \bar \nu_i \quad \propto \quad |y_{33}|^2, \ |y_{32}|^2,\ |y_{33}|^2+|y_{32}|^2 \ , \label{decay2} \\
S_3^{+4/3} &\to& \bar b \tau^+, \ \bar b \mu^+ \quad \propto \quad |y_{33}|^2, \ |y_{32}|^2 \ , \label{decay3}
\end{eqnarray}
and their CP conjugates.  We have indicated, along with each decay mode, the corresponding couplings involved. Here we have neglected all couplings except $y_{33}$ and $y_{32}$, and assumed that the PMNS matrix is unitary and $V_{33}=1$. The exact formulas including the CKM and PMNS mixing matrices are deduced from Eq.~(\ref{eq-LS3-2}).
Note that, although $S_3^{4/3}$ is the component that would be involved in the neutral-current B-physics anomaly $R_K^{(*)}$, its main decay mode ($S_3^{4/3} \to b\tau$) when produced on-shell is hard to detect at the LHC because of difficulties in $\tau$-lepton tagging and large associated backgrounds. Recent dedicated searches \cite{Sirunyan:2017yrk} and constraints from $B-L$ R-parity violating stops~\cite{Aaboud:2017opj}, are used to set limits on this particle.  In contrast, the $S_3^{2/3}$ component of the triplet has a unique decay mode,  $S_3^{2/3} \to t \nu$, which is quite well constrained by stop searches~\cite{Sirunyan:2017kqq}.  On the other hand, $S_3^{1/3}$ is the only component that has decay channels to the upper and lower members of the lepton doublet. The $S_3^{1/3} \to b \nu$ decay mode is constrained by sbottom searches \cite{Sirunyan:2017kqq}, whereas the charged lepton decay mode has limits from dedicated leptoquark searches~\cite{Sirunyan:2018nkj,CMS-PAS-B2G-16-027} and can also be constrained via searches in multi-lepton-plus-jets final states, such as $t\bar{t}t\bar{t}$ (denoted as ``4-top" in the following) searches~\cite{Sirunyan:2017roi}. In this article we focus on the constraints on the $S_3^{1/3}$ leptoquark.

It is interesting to notice in Eqs.~(\ref{decay1}-\ref{decay3}) that these decays are driven by what our framework predicts to be the largest $y_{ij}$. In contrast, because of kinematic considerations, in B-meson decays most of these $y_{3j}$ are not accessible unless they are accompanied by a suppression factor $y_{2j}$, as for instance $y_{22}$ in $B \to K \mu^+ \mu^-$.
In addition to the different phenomenology of these decay channels, we should expect a splitting of the mass eigenstate due to the electroweak symmetry breaking.  Using results in Section \ref{section:2} we can expect a leptoquark mass splitting of $\Delta m_S \sim {\cal O}(100\ \mbox{GeV})$. 

Since the $S_3$ leptoquark is charged under color,  it couples to gluons through the strong interaction (with coupling strength $g_s$) and it can be pair produced at the LHC independently of its weak coupling to quark and leptons.  Moreover, depending on the leptoquark mass and this weak coupling, there will be some characteristic mass limit above which it would be more promising to study single, rather than pair production \cite{Dorsner:2018ynv, Buttazzo:2017ixm}.  Given the current status of the B-physics anomalies, we find that a $S_3$ with couplings motivated by partial compositeness has a larger cross-section for pair production than for single production for leptoquark masses $\lesssim 1-1.5$~TeV (see for instance Ref.~\cite{Marzocca:2018wcf}).  Therefore, in this article we consider solely QCD-mediated leptoquark pair production and leave the investigation of single production within a partial compositeness framework for future work. 

Once leptoquarks are pair produced at the LHC, they decay into quarks and leptons.  The actual branching ratios would be determined by the $y_{ij}$ coefficients, as presented in Eqs.~(\ref{decay1}-\ref{decay3}).  The $R_K^{(*)}$ anomalies suggest the following relationship~\cite{Hiller:2017bzc}:
\begin{equation}
\frac{y_{32} \, y_{22}^* - y_{31} \, y_{21}^*}{M^2} \approx \frac{1}{(33\mbox{ TeV})^2} \ .
\label{mainrelation}
\end{equation}
Since in the context of partial compositeness the couplings involving fermions of the first generation are suppressed compared with those involving the second-generation, Eq.~(\ref{mainrelation}) simplifies to
\begin{equation}
\frac{y_{32}\,  y_{22}^*}{M^2} \approx \frac{1}{(33\mbox{ TeV})^2} \ .
\label{banomaly}
\end{equation}
Within our framework there are further expected relations between the $y_{ij}$ coefficients, which determine a preferred curve in parameter space.  In particular, it is useful to recall the following relationships:
\begin{eqnarray}
\frac{y_{22}}{y_{32}} &\sim & \lambda_C^2 \approx 0.05 \label{rel1} \ ,\\
\frac{y_{33}}{y_{32}} &\sim & \sqrt\frac{m_\tau}{m_\mu} \approx 4 \ , \label{rel2}
\end{eqnarray}
which come from the model description in Section~\ref{section:2}.

Within our framework, with no other decay channel other than $S_3^{1/3} \to t\tau, t \mu, b\nu$, and assuming only $y_{3i}$ couplings, $|V_{33}|=1$, an unitary PMNS matrix and all fermions massless, we can obtain the following branching ratios:
\begin{eqnarray}
\mbox{BR}(S_3^{1/3} \to t \tau)&=&\frac{|y_{33}|^2}{2(|y_{33}|^2+|y_{32}|^2)} \ , \\
\mbox{BR}(S_3^{1/3} \to t \mu)&=&\frac{|y_{32}|^2}{2(|y_{33}|^2+|y_{32}|^2)} \ , \\
\mbox{BR}(S_3^{1/3} \to b \nu)&=&\frac{1}{2} \ . 
\end{eqnarray}
Interestingly, due to CKM and PMNS unitarity and the chosen $SU(2)_{L}$ structure of the interaction Lagrangian, the $\mbox{BR}(S_3^{1/3} \to b \nu)$ is fixed and 
\begin{eqnarray}
\mbox{BR}(S_3^{1/3} \to t \tau)+\mbox{BR}(S_3^{1/3} \to t \mu)&=&\mbox{BR}(S_3^{1/3} \to b \nu)\label{relation} =\frac{1}{2} \ . 
\end{eqnarray}

In a more general model, this relationship may not hold depending on the assumptions made. In models without other decay channels, giving a maximum value for $\mbox{BR}(S_3^{1/3} \to b \nu)$ can set a minimum value for $\mbox{BR}(S_3^{1/3} \to t \tau)+\mbox{BR}(S_3^{1/3}  \to t \mu)$. In our model, it also sets a maximum value to $\mbox{BR}(S_3^{1/3} \to t \tau)+\mbox{BR}(S_3^{1/3}  \to t \mu)$.
By taking Eq.~(\ref{rel2}) and allowing preferred region within half of its central value one obtains for our relevant component $S_3^{1/3}$:
\begin{equation}
\frac{\mbox{BR}(S_3^{1/3} \to t \tau)}{\mbox{BR}(S_3^{1/3} \to t \mu)} = (4 \pm 2)^2 \ .  \label{relation2}
\end{equation}
The size of the preferred region is somewhat arbitrary, it reflects the fact that Eq.~(\ref{rel2}) is determined up to factors of ${\cal O}(1)$.

Equations~(\ref{relation}) and~(\ref{relation2}) define a curve in the branching ratio parameter space in which our pNGB-based leptoquark model is preferred. 
In Section~\ref{section:4} we recast several existing searches to derive bounds on the couplings and mass of  $S_3^{1/3}$, and compare them to the preferred curve within our model. We also discuss the constraints resulting from searches for the other members of the triplet. However, one should bear in mind the possibility of a significant mass splitting, as indicated above.

\section{Direct searches for the $S_3^{1/3}$ leptoquark at the LHC}
\label{section:4}

In a partial compositeness framework --and in many other NP scenarios-- the main decay channels of $S_3^{1/3}$ are expected to be $t\tau$ and $b\nu$, with also a non-negligible contribution from $t\mu$.  
We consider a dedicated search for pair production of $S_3^{1/3}$ decaying into $t\tau$ \cite{Sirunyan:2018nkj}, as well as two recent searches 
in final states with $b\bar{b}+\met$~\cite{Sirunyan:2017kqq}, and multileptons plus $b$-jets (in the context of a 4-top search)~\cite{Sirunyan:2017roi}, 
which would be sensitive to $S_3^{1/3}$.  The CMS Collaboration has recently released a direct search for 
pair production of $S_3^{1/3}$ decaying into $t\mu$~\cite{CMS-PAS-B2G-16-027}, but unfortunately the public document does not contain sufficient 
information for a reinterpretation, and thus could not be considered in our study.
In this section we pay special attention to the 4-top search, not only because it has not been analyzed in this context yet, but also because its translation to the leptoquark parameter space requires a detailed recast of the experimental results. Possible constraints coming from other members of the leptoquark triplet 
are mentioned in Section~\ref{section:5}. Some details of each of the searches considered are provided below:

\begin{enumerate}[i)]

\item {\it Leptoquark search in the $t\bar{t}\tau^+\tau^-$ channel:} 
The CMS Collaboration has recently performed a dedicated leptoquark search in this channel using 35.9 fb$^{-1}$ of $pp$ collision data 
at $\sqrt{s}=13$ TeV~\cite{Sirunyan:2018nkj}. Events are required to have one light lepton ($\ell = e, \mu$), at least one hadronically decaying tau candidate ($\tauhad$), and at least two jets. Different event categories are considered: $\ell$+$\tauhad$ with opposite charge and at least four jets, $\ell$+$\tauhad$ with same charge, and $\ell$+$\geq$2$\tauhad$ (with an opposite-charge $\tauhad$ candidate pair). In the categories with exactly
one $\tauhad$ candidate, the kinematic reconstruction of the top-quark candidate is performed. The main discriminating variable is the $\pt$ of the
top-quark candidate. In contrast, in the category with $\geq$2$\tauhad$ candidates a counting experiment is performed.
Upper limits on $\sigma(pp \to  S_3^{+1/3} S_3^{-1/3}) \times BR(S_3^{1/3} \to t\tau)^2$ are obtained as a function of leptoquark mass ($M_{S_3^{1/3}}$) and compared to the theoretical prediction. For $\BR(S_3^{1/3} \to t \tau)=1$, the 95\% C.L. lower limit on the $S_3^{1/3}$  mass is $M_{S_3^{1/3}}>0.9$ TeV. We recast this result as a function of the different branching ratios considered. Given the experimental requirements summarized above, we make the approximation that the acceptance is similar for signal events with $t\tau t\tau$ and $t\tau t\mu$ 
final states, whereas negligible otherwise.

\item {\it Sbottom search in the $b\bar{b}+\met$ final state:} 
For a massless neutralino, the process $p p \to \tilde b \bar{\tilde b} \to b \bar b \tilde \chi_1^0 \bar \chi_1^0$ has the same final state signature as 
$p p \to S_3^{1/3} \bar S_3^{1/3} \to \bar b \bar \nu b \nu$. 
Therefore, searches for direct sbottom production at the LHC~\cite{Aaboud:2017wqg,Sirunyan:2017kqq} are well suited to probe  the $S_3^{1/3} \to b \nu$ 
decay mode. The most restrictive 95\% CL lower limit on the bottom mass for a massless neutralino, $m_{\tilde b}>1.18$ TeV, is obtained by the 
CMS Collaboration using 35.9 fb$^{-1}$ of $pp$ collision data at $\sqrt{s}=13$ TeV~\cite{Sirunyan:2017kqq}. This limit would be translated to the identical bound for the $S_3^{1/3}$ mass under the hypothetical assumption that BR$(S_3^{1/3} \to b \nu) = 1$, and the value of theory cross section at this mass would then represent the experimental upper limit on the cross section. Considering the leptoquark model in this work, when this branching ratio is reduced because of the presence of new decay modes, which are assumed to have zero acceptance in this analysis, we estimate the new lower mass limit from the crossing of the new theory prediction as a function of mass with the estimated experimental upper limit on the cross section.

\item {\it 4-top search in multilepton final states:}  
As a rare process, a 4-top search is sensitive to many beyond-SM scenarios that involve third-generation quarks.  One of the most powerful signatures to search for 
4-top events involves the presence of multiple leptons with additional $b$-jets, which can also be produced by pair-produced $S_3^{1/3}$ leptoquarks
that decay into $t\tau$ and/or $t\mu$.
The CMS Collaboration has recently performed a search for SM 4-top production in multilepton finals states (only electrons and muons are considered)
using 35.9 fb$^{-1}$ of $pp$ collision data at $\sqrt{s}=13$ TeV~\cite{Sirunyan:2017roi}.
Up to eight signal regions are defined depending on the number of light leptons (same-charge dileptons or trileptons) and the number of $b$-tagged jets 
(2, 3, and $\geq$4). Different requirements on the number of jets are made in different signal regions. In order to recast this search, a detailed study has 
been performed attempting  to reproduce the experimental selection and the statistical analysis.

We have implemented our model using {\tt Feynrules}~\cite{Alloul:2013bka} and loaded it into {\tt Madgraph\;5}~\cite{Alwall:2014hca}, which has been
used to simulate $S_{3}^{1/3}$ pair production at $\sqrt{s}=13$~TeV, followed by the $S_{3}^{1/3}$ decay into both $t\tau$ and $t\mu$.
Since we are focusing on pair production, our results are in principle independent of the absolute value of the leptoquark couplings to quarks and leptons.  
At most, their effects comes from the leptoquark width, which is taken to be 1\% of the mass, and thus smaller than the experimental resolution.
The signal events have been generated using a leading-order matrix element and the NN23LO1 PDF set~\cite{Ball:2012cx}, and have been processed 
through {\tt Pythia\;8}~\cite{Sjostrand:2014zea} for the modeling of parton showering and hadronization, as well as through a simulation 
of the CMS detector response as implemented in {\tt Delphes}~\cite{deFavereau:2013fsa}. 
The simulated signal samples have been normalized using the NLO cross sections computed in Ref.~\cite{Borschensky:2014cia} for
squark pair production, which also apply to the case of scalar leptoquark pair production. 

For a more accurate recasting of this search, we have modified the {\tt Delphes} card and our data analysis to match as closely as possible the CMS 
detector and analysis as described in Ref.~\cite{Sirunyan:2017roi}.  With this modification we obtain in our simulations 4-top yields in each signal region that are in an average 15\% disagreement with the central values reported in Ref.~\cite{Sirunyan:2017roi}.  
We scan the parameter space of branching ratios in our model, allowing for $S_3^{1/3}$ decays into $t\tau$, $t\mu$, and $b\nu$, and obtain the predicted 
yields in each signal region for this analysis. The slight discrepancies that had been found in the validation of 4-top yields are used to correct the yields
per signal region. We derive upper limits on the signal cross section times branching ratio using the CL$_{\textrm{s}}$ method \cite{Read:2002hq}, which 
employs as test statistic the ratio of the likelihoods under the signal plus background hypothesis over the background-only hypothesis. The likelihood fits
are performed using the \texttt{HistFitter} package \cite{Baak:2014wma}, which relies on \texttt{RooFit} \cite{RooFit:manual} and the minimization algorithms from \texttt{MINUIT} \cite{James:1994vla}.
\end{enumerate}

\section{Results}
\label{section:5}

In this section we use the results from the LHC searches discussed in Sect.~\ref{section:4} to derive bounds in the plane of 
$\BR(S_3^{1/3} \to t \tau)$ vs $\BR(S_3^{1/3} \to t \mu)$ as a function of $S_3^{1/3}$ mass, under the assumption that 
$\BR(S_3^{1/3} \to t \tau)+\BR(S_3^{1/3} \to t \mu)+\BR(S_3^{1/3} \to b \nu)=1$. We also perform a simple extrapolation of
these searches to higher energy ($\sqrt{s}=14$ TeV) and luminosity ($300\ $fb$^{-1}$), representative of the full LHC Run 3 dataset.
These bounds are compared to the preferred region within our model, given by Eqs.~(\ref{relation}) and~(\ref{relation2}) .

\begin{figure}[p]
\begin{center}
\subfloat[]{\includegraphics[width=0.45\textwidth]{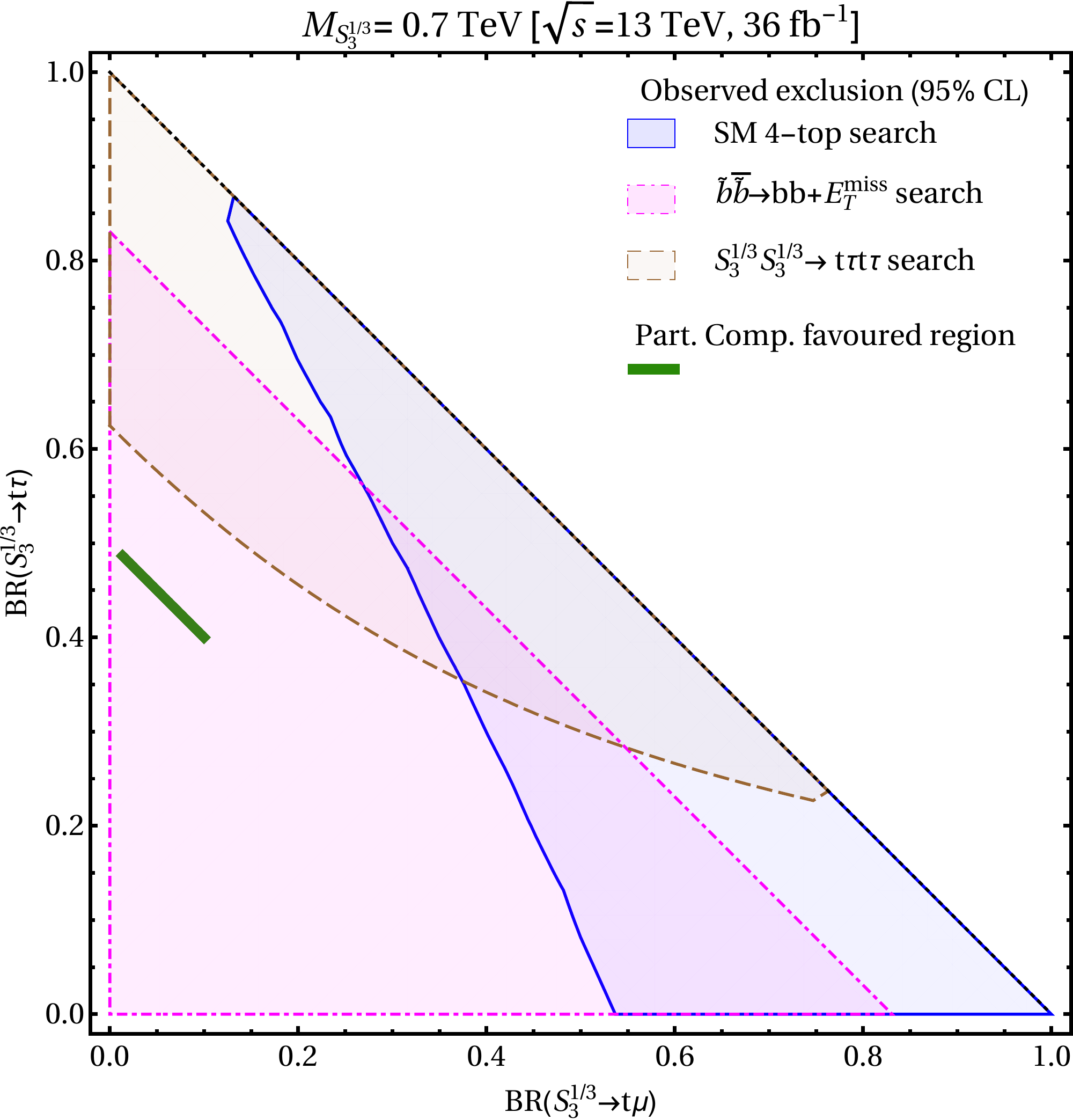}}\hspace{3mm}
\subfloat[]{\includegraphics[width=0.45\textwidth]{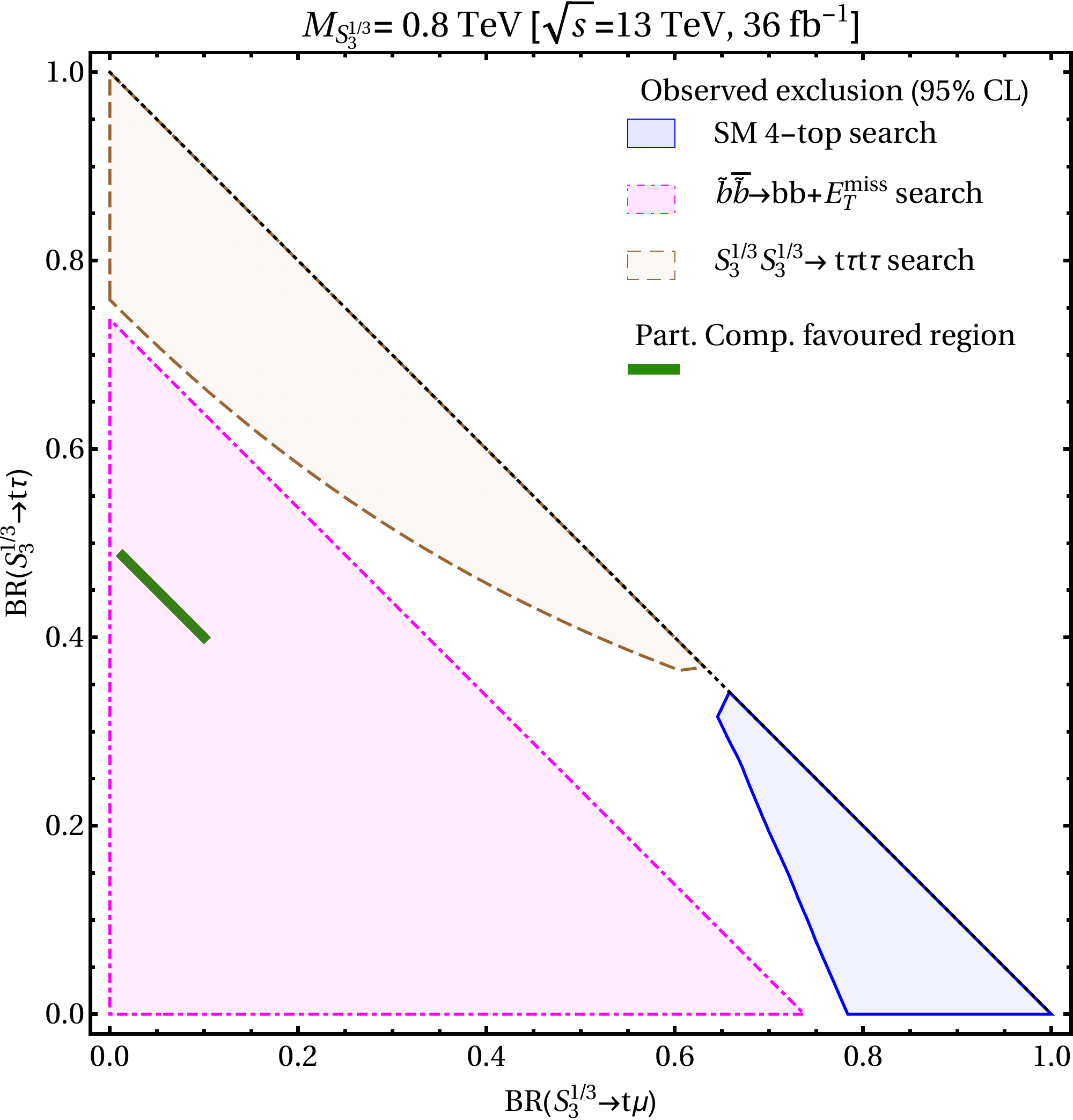}} \\
\subfloat[]{\includegraphics[width=0.45\textwidth]{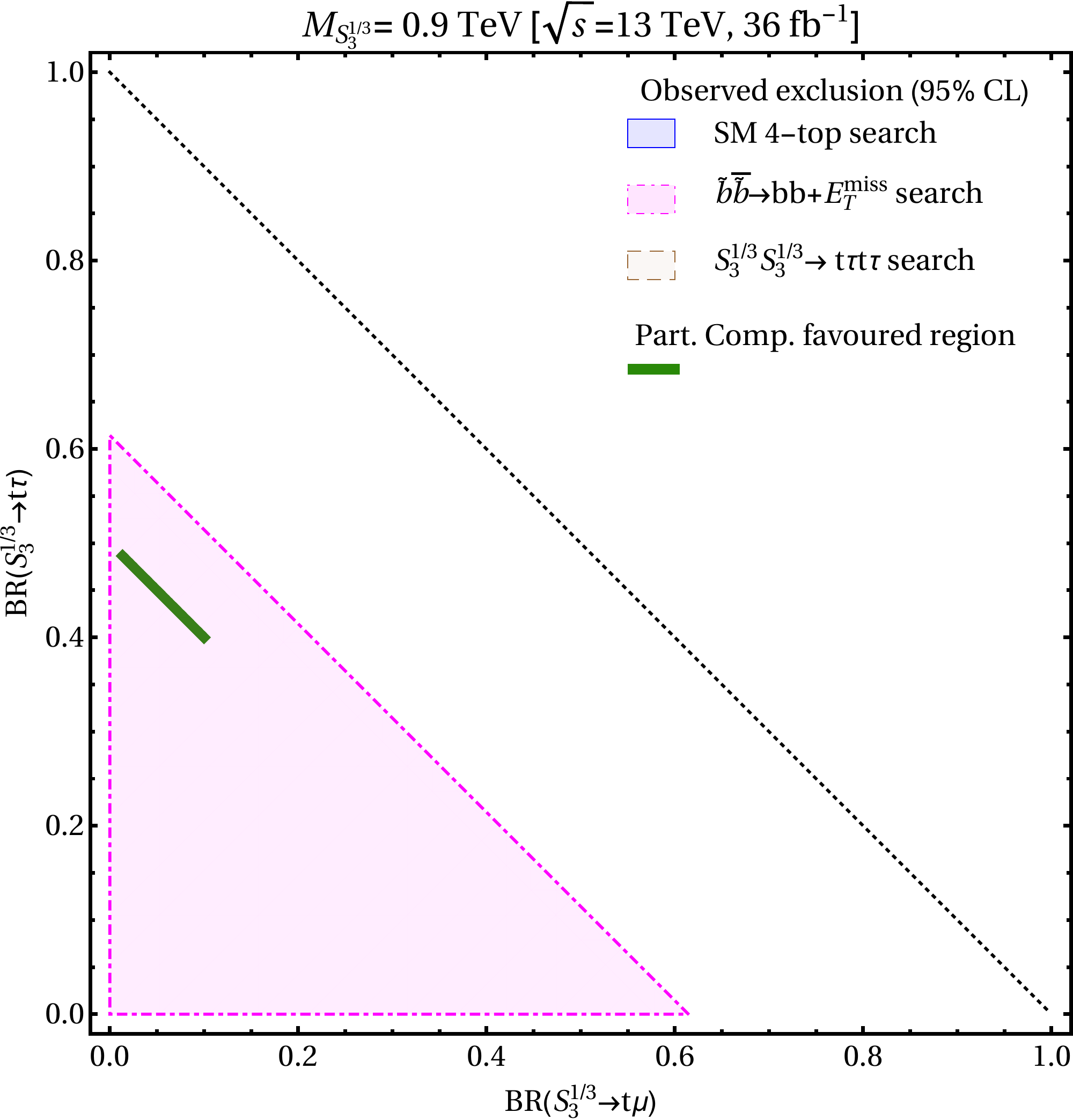}}\hspace{3mm}
\subfloat[]{\includegraphics[width=0.45\textwidth]{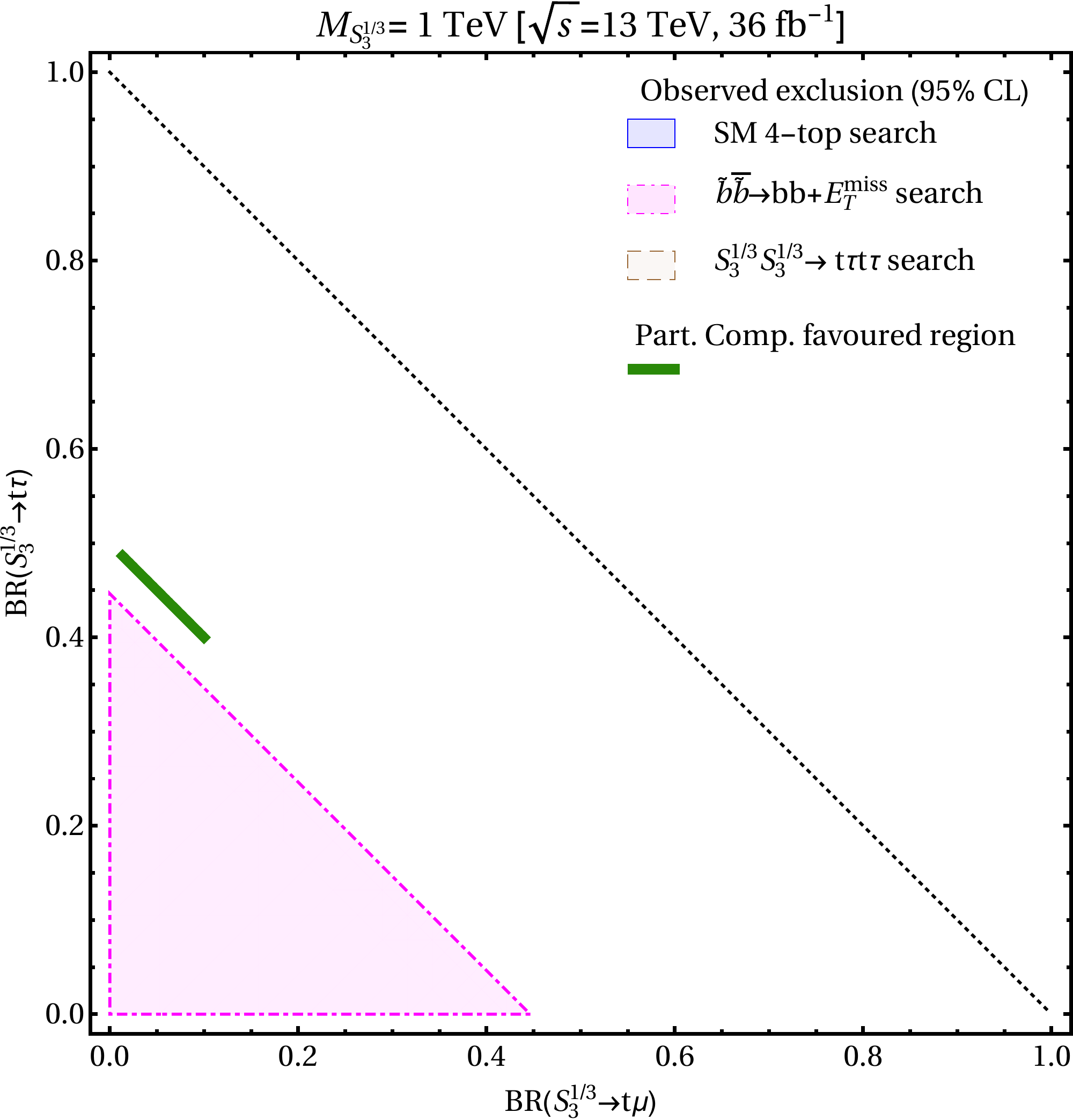}} \\
\caption{\small Bounds on the $S_3^{1/3}$ leptoquark branching ratios into $t\tau$, $t\mu$, and $b\nu$, derived from recent LHC searches based on 
36 fb$^{-1}$ of $pp$ collisions at $\sqrt{s}=13$ TeV, for different values of the leptoquark mass $M_{S_3^{1/3}}$:
(a) 0.7 TeV, (b) 0.8 TeV, (c) 0.9 TeV, and (d) 1 TeV. Also shown is the region preferred by our partial compositeness model.
The area above the diagonal line corresponds to the unphysical region where the sum of branching ratios exceeds unity, or is smaller than zero.}
\label{bounds}
\end{center}
\end{figure}

The three LHC searches that we study probe different final states and thus exhibit complementary sensitivity to the branching ratio parameter space,
as shown in Fig.~\ref{bounds} for different values of $M_{S_3^{1/3}}$.
The direct leptoquark search \cite{Sirunyan:2018nkj} explores the $t\bar{t}\tau^+\tau^-$ final state and thus is sensitive to large 
$\BR(S_3^{1/3} \to t \tau)$, while the sbottom search \cite{Sirunyan:2017kqq}, focused on the $b\bar{b}+\met$ final state, probes 
large $\BR(S_3^{1/3} \to b \nu)$. In the framework of partial compositeness these represent the two dominant decay modes for the $S_3^{1/3}$ leptoquark.
In contrast, the multilepton final states covered by the 4-top search \cite{Sirunyan:2017roi} can originate from $t\bar{t}\mu^+\mu^-$, 
$t\bar{t}\tau^\pm\mu^\mp$, and $t\bar{t}\tau^+\tau^-$. Therefore, the 4-top search provides unique sensitivity to a large generation mixing $y_{32}$ which, 
although is not preferred by partial compositeness, is one of the parameters directly involved in the B-physics anomalies.
It is worth recalling that this presentation of the results, where the bottom left corner of the branching ratio plane corresponds to 
large $\BR(S_3^{1/3} \to b \nu)$, relies on the assumption of only three sizable decay modes ($S_3^{1/3} \to t \tau, t\mu, b\nu$), and would be
modified if additional sizable unknown decay modes would exist. Furthermore, it should be stressed that the parameter space being probed is a 
function of the branching ratios and not the absolute values of the weak couplings. Therefore, since $M_{S_3^{1/3}}>> m_t$ yields branching ratios independent of $M_{S_3^{1/3}}$, practically the whole parameter space as function of branching ratios in Fig.~\ref{bounds} is in principle allowed by the B-physics anomalies, namely Eq.~(\ref{mainrelation}). (The extreme cases of branching ratios equal to zero, one, or very close to them, may yield $y_{ij}$ that either cannot satisfy Eq.~(\ref{mainrelation}) or have non-perturbative values.)

Based on these results, the partial compositeness preferred region is excluded for masses $M_{S_3^{1/3}} < 0.9$ TeV.  
For higher masses, the region preferred by our model is not directly excluded. This conclusion assumes that searches for the other components 
of the multiplet --namely $S_3^{4/3}$ and $S_3^{2/3}$-- do not affect the $S_3^{1/3}$ component. This depends on the mass splitting 
$\Delta m_S\sim {\cal O}(100\ \mbox{GeV})$ and the sensitivity of these other searches. In particular, the $S_3^{-2/3}$ leptoquark decays with 100\% into $t\nu$ and 
thus is constrained by stop searches in the $t\bar{t}+\met$ final state, which yield a mass limit $M_{S_3^{2/3}} > 1.07$ TeV~\cite{Sirunyan:2017kqq}. In contrast, the $S_3^{4/3}$ leptoquark can be sought directly \cite{Sirunyan:2017yrk} or probed by the $B-L$ $R$-parity-violating stop search in Ref.~\cite{Aaboud:2017opj}. 
Please note that the final state in this search is the same as in $S_3^{4/3}$ pair-production except for a bottom-antibottom distinction, 
which becomes irrelevant at the detector level. The actual lower mass limit on $M_{S_3^{4/3}}$ depends on its branching ratios: 
for large branching ratio into $b\tau$ the limit is $\sim$0.9 TeV, whereas for large branching ratio into $b\mu$ the limit reaches $\sim$1.4 TeV.

It is interesting to project the bounds on the $S_3^{1/3}$ parameter space for upcoming LHC runs.
We use the following procedure to estimate the projected sensitivities for the full Run 3 dataset ($\sqrt{s}=14$ TeV, 300 fb$^{-1}$) of the relevant searches 
discussed above.  We obtain the signal cross-section at $\sqrt{s}=14$ TeV from the NLO calculation for squark pair production in Ref.~\cite{Borschensky:2014cia}.
For the direct leptoquark search in the $t\bar{t}\tau^+\tau^-$ channel we assume that background yields scale with a common factor of 1.2 going from 
$\sqrt{s}=13$ TeV to 14 TeV, and that the analysis is statistically limited and thus background uncertainties scale accordingly with the integrated luminosity. 
For the sbottom search we use a similar procedure and we obtain results consistent with those in Ref.~\cite{Ruhr:2016xsg}.  
To project the limits from the 4-top search we rescale the estimated backgrounds to their NLO cross section at $\sqrt{s}=14$ TeV obtained from 
{\tt MadGraph5\_aMC@NLO}~\cite{Alwall:2014hca}, and we perform the statistical analysis based on the expected signal and background yields.

\begin{figure}[p]
\begin{center}
\subfloat[]{\includegraphics[width=0.47\textwidth]{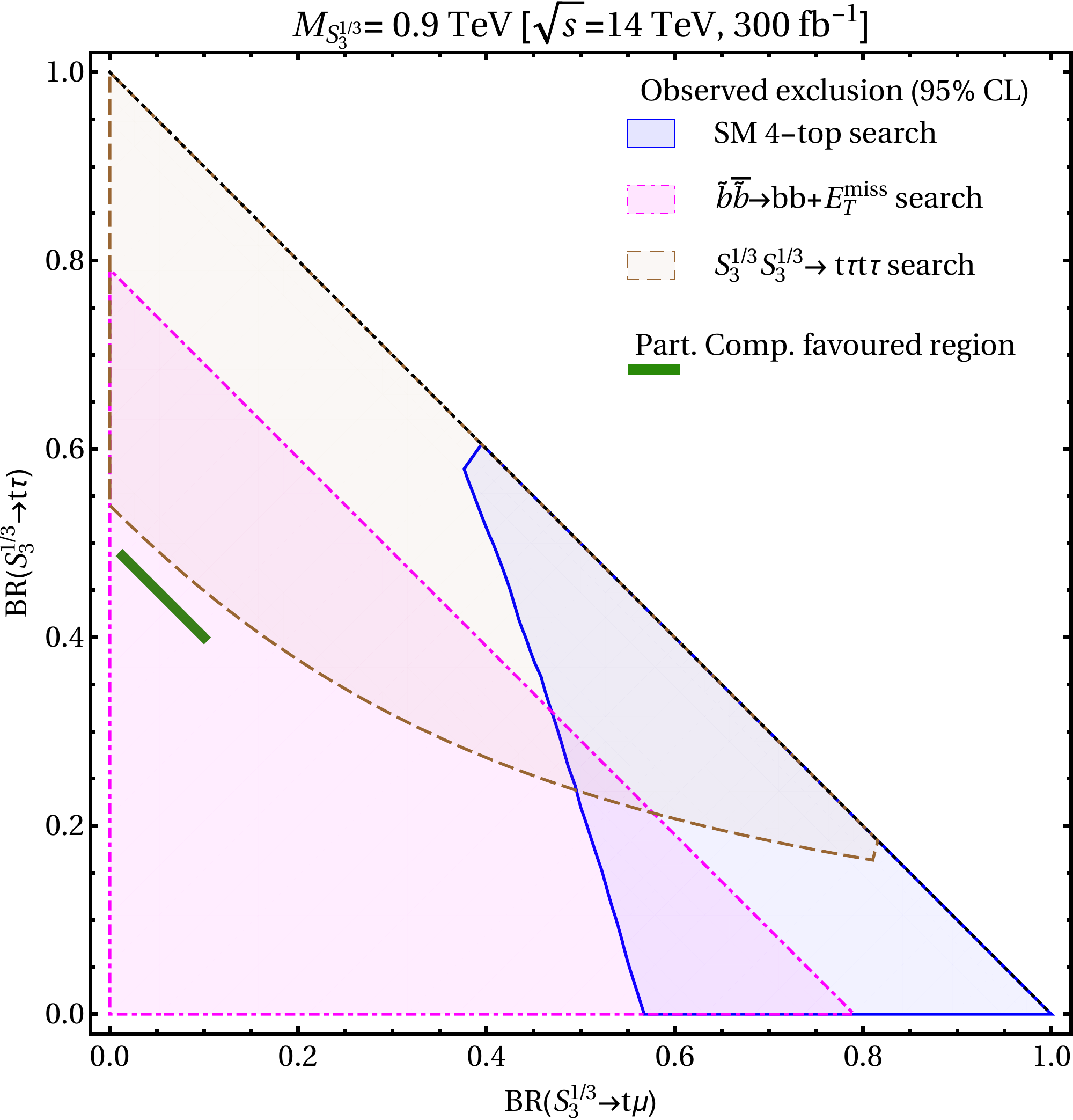}}\hspace{3mm}
\subfloat[]{\includegraphics[width=0.47\textwidth]{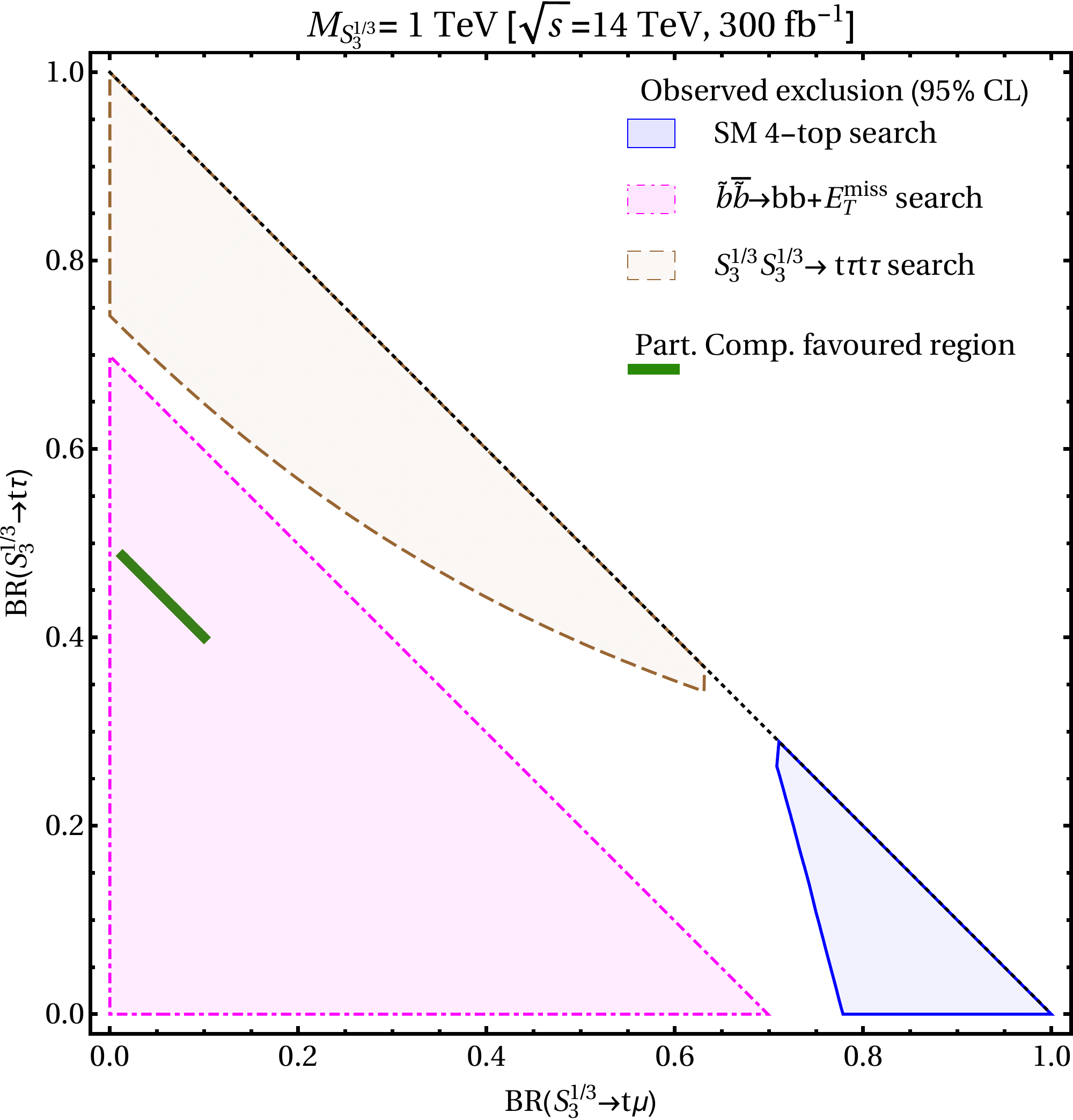}} \\
\subfloat[]{\includegraphics[width=0.47\textwidth]{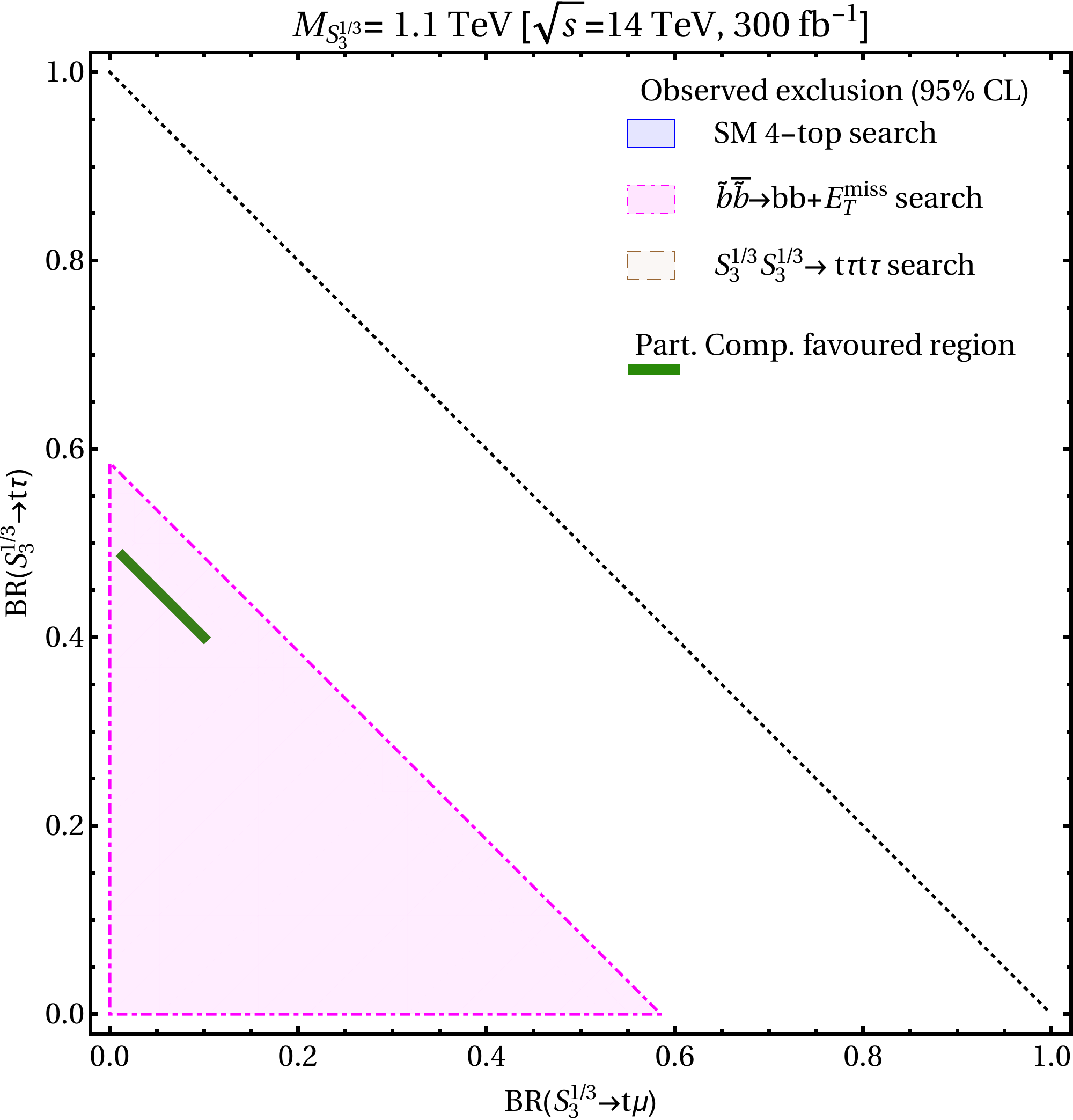}}
\caption{\small Projected bounds on the $S_3^{1/3}$ leptoquark branching ratios into $t\tau$, $t\mu$, and $b\nu$ based on recent LHC searches extrapolated to 
300 fb$^{-1}$ of $pp$ collisions at $\sqrt{s}=14$ TeV, for different values of the leptoquark mass $M_{S_3^{1/3}}$:
(a) 0.9 TeV, (b) 1 TeV, and (c) 1.1 TeV. Also shown is the region preferred by our partial compositeness model.
The area above the diagonal line corresponds to the unphysical region where the sum of branching ratios exceeds unity, or is smaller than zero.}
\label{projected}
\end{center}
\end{figure}

\begin{figure}[t]
\begin{center}
\includegraphics[width=0.6\textwidth]{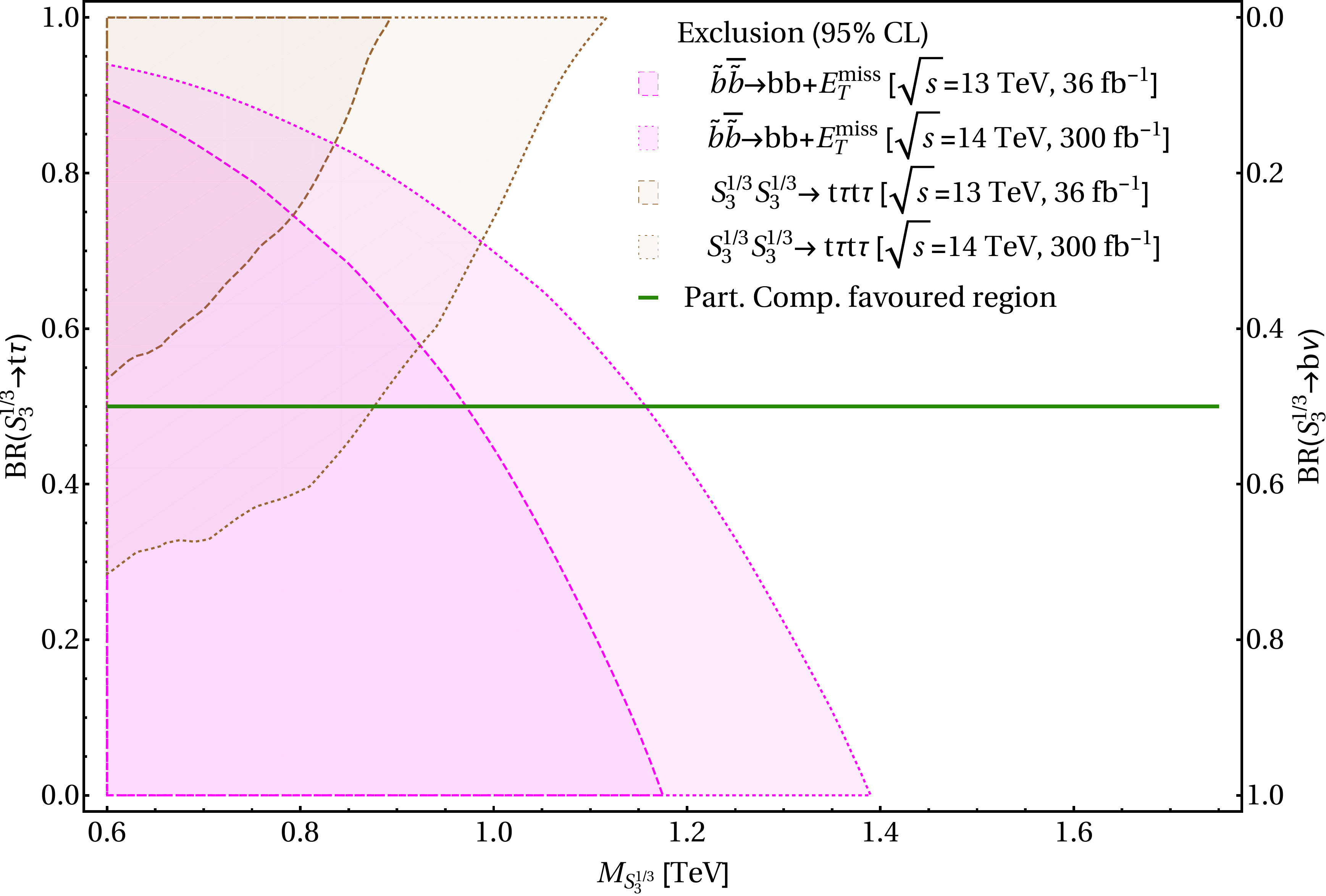}
\caption{ \small Bounds on the $S_3^{1/3}$ leptoquark branching ratios into $t\tau$ and $b\nu$ (assuming a negligible $t\mu$ branching ratio) 
as a function of leptoquark mass $M_{S_3^{1/3}}$. The current bounds (dashed line) and projected bounds (dotted line) for the LHC Run 3 derived
from the $t\bar{t}\tau^+\tau^-$ and $b\bar{b}+\met$ searches are compared.
Also shown is the region preferred by our partial compositeness model.}
\label{moneyplot}
\end{center}
\end{figure}

The projected exclusion regions are shown in Fig.~\ref{projected}, which can be compared to the current ones in Fig.~\ref{bounds}. 
Using the full Run 3 dataset, the full branching ratio plane would be excluded for masses $\leq 0.9$~TeV and a significant portion of the
parameter space can be probed for a mass of $\sim 1$~TeV. 
To better visualize the LHC reach we display in Fig.~\ref{moneyplot} the exclusion regions under the assumption that 
$\BR(S_3^{1/3} \to t \mu)$ is negligible, as expected in partial compositeness. In this case we consider only the searches in the
$t\bar{t}\tau^+\tau^-$ and $b\bar{b}+\met$ final states. For the scenario $\BR(S_3^{1/3} \to t \tau) = \BR(S_3^{1/3} \to b \nu) = 0.5$, 
masses up to 1.2 TeV can be probed, driven by the $b\bar{b}+\met$ search. A dedicated search targeting the $t\tau b\nu$ final state, which would have the
largest branching ratio, is potentially interesting to probe masses somewhat higher than 1.2 TeV.  In any case, exploring beyond the 1--1.5 TeV mass scale 
for the $S_3^{1/3}$ leptoquark will require pursuing different strategies, such as considering the singly-resonant and non-resonant production mechanisms.

\section{Summary and outlook}
\label{section:6}

In this paper we use the results from recent LHC searches to constrain the parameter space of a model with a scalar leptoquark ($S_3$) 
that transforms as $(\bar 3, 3)_{1/3}$.  This model is motivated by the B-physics anomalies and encloses the $S_3$ state and the SM Higgs 
boson as pNGBs of a strongly interacting sector at higher energies. The pNGB scenario allows to justify why this scalar particle would be 
the lightest new state in the spectrum and thus potentially within reach at the LHC, whereas other fermionic and vector resonances
would be heavier and potentially beyond kinematic reach.  We also exploit the relationships between couplings of the leptoquark to quarks 
and leptons of different generations that arise within the framework of partial compositeness.

We study the production of a pair of leptoquarks, $pp\to S_3 \bar S_3$, since its cross section does not depend on its couplings
to leptons and quarks,  and it should represent the dominant production mechanism for a leptoquark in the mass regime currently probed
at the LHC (up to $\sim$~1 TeV). The different decay modes by each of the three $S_3$ components result in different search strategies.
The $S_3^{2/3}$ leptoquark is the most straightforward one to probe since its three decay channels all lead 
to the same final state signature, $S_3^{-2/3}S_3^{+2/3} \to t\bar{t}+\met$.
Consequently, current stop searches can be effectively used to exclude $S_3^{-2/3}$ masses below $\sim$1.1~GeV~\cite{Sirunyan:2017kqq}.
In contrast, pair production of the $S_3^{4/3}$ leptoquark results in final states with pairs of opposite-charge leptons plus $b$-quark jets,
$S_3^{-4/3} S_3^{+4/3} \to b\ell \bar{b} \ell'$  ($\ell, \ell' = e, \mu, \tau$), although there is a prejudice towards same-flavor dilepton final states
with the lepton being $\mu$ or $\tau$.
Lower mass limits for this leptoquark come from dedicated searches \cite{Sirunyan:2017yrk} and from constraints in $B-L$ $R$-parity violating stop searches \cite{Aaboud:2017opj} and, depending on the assumed branching ratio, can range from $\sim$0.9~GeV (mainly $b\tau$) to $\sim$1.4~GeV (mainly $b\mu$).
Finally, the $S_3^{1/3}$ leptoquark has six decay channels, giving multiple final states that can be probed at the LHC: 
$S_3^{+1/3} S_3^{-1/3} \to t\ell \bar{t} \ell'$,  $b\bar{b} + \met$, and $t \ell b + \met$. Therefore, the LHC is in principle more sensitive to probe this leptoquark
Since we can expect within our model a mass splitting between the components of the $S_3$ multiplet of up to about ${\cal O} (100\mbox{ GeV})$, 
lower mass limits derived for one component cannot a-priori be applied to other components.  In this work we have mainly focused on constraining the parameter 
space of the $S_3^{1/3}$ component.  Further work within the proposed model could relate these constraints to limits in the other components.

Within our model, the main decay channels of the $S_3^{1/3}$ leptoquark are $S_3^{1/3} \to t\tau, t \mu, b\nu$.
We study three complementary searches at the LHC using 36 fb$^{-1}$ at $\sqrt{s}=13$~TeV that can constrain the corresponding branching 
ratio parameter space.
These include dedicated searches for $t\bar{t}\tau^+\tau^-$~\cite{Sirunyan:2018nkj} and $b\bar{b}+\met$~\cite{Aaboud:2017wqg,Sirunyan:2017kqq}, 
which are primarily sensitive to large $\BR(S_3^{1/3} \to t \tau)$ and $\BR(S_3^{1/3} \to b \nu)$, respectively. In addition, we recast a search for 
SM 4-top production in multilepton final states~\cite{Sirunyan:2017roi}, which is sensitive to large $\BR(S_3^{1/3} \to t \mu)$.
We find that most of the branching ratio parameter space is excluded up to masses of about $\sim$800 GeV, and that the preferred region for
our model, $\BR(S_3^{1/3} \to t \tau) \simeq \BR(S_3^{1/3} \to b \nu) \simeq 0.5$, is best probed by the $b\bar{b}+\met$ search and remains allowed for
masses above 1~TeV. 
We also project the sensitivity of these searches to the full LHC Run 3 dataset (300 fb$^{-1}$ at $\sqrt{s}=14$~TeV).
We find that for a dominant branching ratio to $t\tau$ ($b\nu$) it will be possible to exclude masses up to 1.1~TeV (1.4~TeV), whereas for the more
realistic scenario of equally split branching ratios, as predicted by our model, the expected limit is about 1.2~TeV.  
In this case, a dedicated search targeting the $t\tau b\nu$ final state would be well motivated to further extend the sensitivity.

Finally, we appraise that for mass scales well above 1--1.5~TeV, and for regions of parameter space consistent with the B-physics anomalies,
singly-resonant and non-resonant production are expected to dominate, requiring optimized search strategies.
We leave this investigation for future work.

\section*{Acknowledgments}

E.A., L.D.~and M.S.'s~work is partially supported by ANPCyT PICT 2013-2266.
A.J. is supported in part by the Spanish Ministerio de Econom\'ia y Competitividad  under projects
FPA2015-69260-C3-1-R and Centro de Excelencia Severo Ochoa SEV-2012-0234.

\bibliographystyle{JHEP}
\bibliography{biblio}

\providecommand{\href}[2]{#2}\begingroup\raggedright\begin{thebibliography}{10}

\bibitem{Aaij:2014ora}
{LHCb Collaboration}, \emph{{Test of lepton universality using
  $B^{+}\rightarrow K^{+}\ell^{+}\ell^{-}$ decays}},
  \href{http://dx.doi.org/10.1103/PhysRevLett.113.151601}{\emph{Phys. Rev.
  Lett.} {\bf 113} (2014) 151601},
  [\href{http://arxiv.org/abs/arXiv:1406.6482}{{\tt arXiv:1406.6482}}].

\bibitem{Aaij:2015yra}
{LHCb Collaboration}, \emph{{Measurement of the ratio of branching fractions
  $\mathcal{B}(\bar{B}^0 \to
  D^{*+}\tau^{-}\bar{\nu}_{\tau})/\mathcal{B}(\bar{B}^0 \to
  D^{*+}\mu^{-}\bar{\nu}_{\mu})$}},
  \href{http://dx.doi.org/10.1103/PhysRevLett.115.159901,
  10.1103/PhysRevLett.115.111803}{\emph{Phys. Rev. Lett.} {\bf 115} (2015)
  111803}, [\href{http://arxiv.org/abs/arXiv:1506.08614}{{\tt
  arXiv:1506.08614}}]. [Erratum: Phys. Rev. Lett. 115 (2015) 159901].

\bibitem{Huschle:2015rga}
{Belle Collaboration}, \emph{{Measurement of the branching ratio of $\bar{B}
  \to D^{(\ast)} \tau^- \bar{\nu}_\tau$ relative to $\bar{B} \to D^{(\ast)}
  \ell^- \bar{\nu}_\ell$ decays with hadronic tagging at Belle}},
  \href{http://dx.doi.org/10.1103/PhysRevD.92.072014}{\emph{Phys. Rev. D} {\bf
  92} (2015) 072014}, [\href{http://arxiv.org/abs/arXiv:1507.03233}{{\tt
  arXiv:1507.03233}}].

\bibitem{Aaij:2017vbb}
{LHCb Collaboration}, \emph{{Test of lepton universality with $B^{0}
  \rightarrow K^{*0}\ell^{+}\ell^{-}$ decays}},
  \href{http://dx.doi.org/10.1007/JHEP08(2017)055}{\emph{JHEP} {\bf 08} (2017)
  055}, [\href{http://arxiv.org/abs/arXiv:1705.05802}{{\tt arXiv:1705.05802}}].

\bibitem{Buttazzo:2017ixm}
D.~Buttazzo, A.~Greljo, G.~Isidori and D.~Marzocca, \emph{{B-physics anomalies:
  a guide to combined explanations}},
  \href{http://dx.doi.org/10.1007/JHEP11(2017)044}{\emph{JHEP} {\bf 11} (2017)
  044}, [\href{http://arxiv.org/abs/arXiv:1706.07808}{{\tt arXiv:1706.07808}}].

\bibitem{Hiller:2017bzc}
G.~Hiller and I.~Nisandzic, \emph{{$R_K$ and $R_{K^{\ast}}$ beyond the standard
  model}}, \href{http://dx.doi.org/10.1103/PhysRevD.96.035003}{\emph{Phys. Rev.
  D} {\bf 96} (2017) 035003},
  [\href{http://arxiv.org/abs/arXiv:1704.05444}{{\tt arXiv:1704.05444}}].

\bibitem{Calibbi:2015kma}
L.~Calibbi, A.~Crivellin and T.~Ota, \emph{{Effective Field Theory Approach to
  $b\to s \ell\ell^{(\prime)}$, $B\to K^{(*)}\nu\bar{\nu}$ and $B\to
  D^{(*)}\tau\nu$ with Third Generation Couplings}},
  \href{http://dx.doi.org/10.1103/PhysRevLett.115.181801}{\emph{Phys. Rev.
  Lett.} {\bf 115} (2015) 181801},
  [\href{http://arxiv.org/abs/arXiv:1506.02661}{{\tt arXiv:1506.02661}}].

\bibitem{Becirevic:2015asa}
D.~Be\v{c}irevi\'c, S.~Fajfer and N.~Ko\v{s}nik, \emph{{Lepton flavor
  nonuniversality in $b\to s\ell^+\ell^-$ processes}},
  \href{http://dx.doi.org/10.1103/PhysRevD.92.014016}{\emph{Phys. Rev. D} {\bf
  92} (2015) 014016}, [\href{http://arxiv.org/abs/arXiv:1503.09024}{{\tt
  arXiv:1503.09024}}].

\bibitem{Dorsner:2017ufx}
I.~Doršner, S.~Fajfer, D.~A. Faroughy and N.~Košnik, \emph{{The role of the
  $S_3$ GUT leptoquark in flavor universality and collider searches}},
  \href{http://dx.doi.org/10.1007/JHEP10(2017)188}{\emph{JHEP} {\bf 10} (2017)
  188}, [\href{http://arxiv.org/abs/arXiv:1706.07779}{{\tt arXiv:1706.07779}}].

\bibitem{Crivellin:2017zlb}
A.~Crivellin, D.~Muller and T.~Ota, \emph{{Simultaneous explanation of
  R($D^{(*)}$) and $b\rightarrow s\mu^{+}\mu^{-}$: the last scalar leptoquarks
  standing}}, \href{http://dx.doi.org/10.1007/JHEP09(2017)040}{\emph{JHEP} {\bf
  09} (2017) 040}, [\href{http://arxiv.org/abs/arXiv:1703.09226}{{\tt
  arXiv:1703.09226}}].

\bibitem{Pati:1973uk}
J.~C. Pati and A.~Salam, \emph{{Unified Lepton-Hadron Symmetry and a Gauge
  Theory of the Basic Interactions}},
  \href{http://dx.doi.org/10.1103/PhysRevD.8.1240}{\emph{Phys. Rev. D} {\bf 8}
  (1973) 1240}.

\bibitem{Becirevic:2016yqi}
D.~Bečirević, S.~Fajfer, N.~Ko\v{s}nik and O.~Sumensari, \emph{{Leptoquark
  model to explain the $B$-physics anomalies, $R_K$ and $R_D$}},
  \href{http://dx.doi.org/10.1103/PhysRevD.94.115021}{\emph{Phys. Rev. D} {\bf
  94} (2016) 115021}, [\href{http://arxiv.org/abs/arXiv:1608.08501}{{\tt
  arXiv:1608.08501}}].

\bibitem{Kosnik:2012dj}
N.~Ko\v{s}nik, \emph{{Model independent constraints on leptoquarks from $b \to
  s \ell^+ \ell^-$ processes}},
  \href{http://dx.doi.org/10.1103/PhysRevD.86.055004}{\emph{Phys. Rev. D} {\bf
  86} (2012) 055004}, [\href{http://arxiv.org/abs/arXiv:1206.2970}{{\tt
  arXiv:1206.2970}}].

\bibitem{Fajfer:2017lzq}
S.~Fajfer, \emph{{Scalar (or vector) leptoquarks in B meson anomalies}},
  \href{http://dx.doi.org/10.1016/j.nuclphysbps.2017.03.015}{\emph{Nucl. Part.
  Phys. Proc.} {\bf 285} (2017) 81}.

\bibitem{Angelescu:2018tyl}
A.~Angelescu, D.~Bečirević, D.~A. Faroughy and O.~Sumensari, \emph{{Closing
  the window on single leptoquark solutions to the $B$-physics anomalies}},
  \href{http://arxiv.org/abs/1808.08179}{{\tt 1808.08179}}.

\bibitem{Becirevic:2018afm}
D.~Bečirević, I.~Doršner, S.~Fajfer, N.~Košnik, D.~A. Faroughy and
  O.~Sumensari, \emph{{Scalar leptoquarks from grand unified theories to
  accommodate the $B$-physics anomalies}},
  \href{http://dx.doi.org/10.1103/PhysRevD.98.055003}{\emph{Phys. Rev. D} {\bf
  98} (2018) 055003}, [\href{http://arxiv.org/abs/1806.05689}{{\tt
  1806.05689}}].

\bibitem{Marzocca:2018wcf}
D.~Marzocca, \emph{{Addressing the B-physics anomalies in a fundamental
  Composite Higgs Model}},
  \href{http://dx.doi.org/10.1007/JHEP07(2018)121}{\emph{JHEP} {\bf 07} (2018)
  121}, [\href{http://arxiv.org/abs/arXiv:1803.10972}{{\tt arXiv:1803.10972}}].

\bibitem{Gripaios:2009dq}
B.~Gripaios, \emph{{Composite Leptoquarks at the LHC}},
  \href{http://dx.doi.org/10.1007/JHEP02(2010)045}{\emph{JHEP} {\bf 02} (2010)
  045}, [\href{http://arxiv.org/abs/arXiv:0910.1789}{{\tt arXiv:0910.1789}}].

\bibitem{Gripaios:2014tna}
B.~Gripaios, M.~Nardecchia and S.~A. Renner, \emph{{Composite leptoquarks and
  anomalies in $B$-meson decays}},
  \href{http://dx.doi.org/10.1007/JHEP05(2015)006}{\emph{JHEP} {\bf 05} (2015)
  006}, [\href{http://arxiv.org/abs/arXiv:1412.1791}{{\tt arXiv:1412.1791}}].

\bibitem{leandro}
L.~D. Rold and F.~Lamagna, \emph{{In preparation}}, .

\bibitem{Gherghetta:2000qt}
T.~Gherghetta and A.~Pomarol, \emph{{Bulk fields and supersymmetry in a slice
  of AdS}}, \href{http://dx.doi.org/10.1016/S0550-3213(00)00392-8}{\emph{Nucl.
  Phys. B} {\bf 586} (2000) 141},
  [\href{http://arxiv.org/abs/arXiv:hep-ph/0003129}{{\tt
  arXiv:hep-ph/0003129}}].

\bibitem{Csaki:2008zd}
C.~Csaki, A.~Falkowski and A.~Weiler, \emph{{The Flavor of the Composite
  Pseudo-Goldstone Higgs}},
  \href{http://dx.doi.org/10.1088/1126-6708/2008/09/008}{\emph{JHEP} {\bf 09}
  (2008) 008}, [\href{http://arxiv.org/abs/arXiv:0804.1954}{{\tt
  arXiv:0804.1954}}].

\bibitem{Giudice:2007fh}
G.~F. Giudice, C.~Grojean, A.~Pomarol and R.~Rattazzi, \emph{{The
  Strongly-Interacting Light Higgs}},
  \href{http://dx.doi.org/10.1088/1126-6708/2007/06/045}{\emph{JHEP} {\bf 06}
  (2007) 045}, [\href{http://arxiv.org/abs/arXiv:hep-ph/0703164}{{\tt
  arXiv:hep-ph/0703164}}].

\bibitem{Kaplan:1991dc}
D.~B. Kaplan, \emph{{Flavor at SSC energies: A New mechanism for dynamically
  generated fermion masses}},
  \href{http://dx.doi.org/10.1016/S0550-3213(05)80021-5}{\emph{Nucl. Phys. B}
  {\bf 365} (1991) 259}.

\bibitem{KerenZur:2012fr}
B.~Keren-Zur, P.~Lodone, M.~Nardecchia, D.~Pappadopulo, R.~Rattazzi and
  L.~Vecchi, \emph{{On Partial Compositeness and the CP asymmetry in charm
  decays}},
  \href{http://dx.doi.org/10.1016/j.nuclphysb.2012.10.012}{\emph{Nucl. Phys. B}
  {\bf 867} (2013) 394}, [\href{http://arxiv.org/abs/arXiv:1205.5803}{{\tt
  arXiv:1205.5803}}].

\bibitem{Bauer:2011ah}
M.~Bauer, R.~Malm and M.~Neubert, \emph{{A Solution to the Flavor Problem of
  Warped Extra-Dimension Models}},
  \href{http://dx.doi.org/10.1103/PhysRevLett.108.081603}{\emph{Phys. Rev.
  Lett.} {\bf 108} (2012) 081603},
  [\href{http://arxiv.org/abs/arXiv:1110.0471}{{\tt arXiv:1110.0471}}].

\bibitem{DaRold:2017dbr}
L.~Da~Rold and I.~A. Davidovich, \emph{{A symmetry for $\epsilon_K$}},
  \href{http://dx.doi.org/10.1007/JHEP10(2017)135}{\emph{JHEP} {\bf 10} (2017)
  135}, [\href{http://arxiv.org/abs/arXiv:1704.08704}{{\tt arXiv:1704.08704}}].

\bibitem{DaRold:2017xdm}
L.~Da~Rold, \emph{{Anarchy with linear and bilinear interactions}},
  \href{http://dx.doi.org/10.1007/JHEP10(2017)120}{\emph{JHEP} {\bf 10} (2017)
  120}, [\href{http://arxiv.org/abs/arXiv:1708.08515}{{\tt arXiv:1708.08515}}].

\bibitem{Panico:2015jxa}
G.~Panico and A.~Wulzer, \emph{{The Composite Nambu-Goldstone Higgs}},
  \href{http://dx.doi.org/10.1007/978-3-319-22617-0}{\emph{Lect. Notes Phys.}
  {\bf 913} (2016) pp.1--316},
  [\href{http://arxiv.org/abs/arXiv:1506.01961}{{\tt arXiv:1506.01961}}].

\bibitem{Dorsner:2016wpm}
I.~Doršner, S.~Fajfer, A.~Greljo, J.~F. Kamenik and N.~Ko\v{s}nik,
  \emph{{Physics of leptoquarks in precision experiments and at particle
  colliders}},
  \href{http://dx.doi.org/10.1016/j.physrep.2016.06.001}{\emph{Phys. Rept.}
  {\bf 641} (2016) 1}, [\href{http://arxiv.org/abs/arXiv:1603.04993}{{\tt
  arXiv:1603.04993}}].

\bibitem{Sirunyan:2017yrk}
{\scshape CMS} collaboration, A.~M. Sirunyan et~al., \emph{{Search for
  third-generation scalar leptoquarks and heavy right-handed neutrinos in final
  states with two tau leptons and two jets in proton-proton collisions at $
  \sqrt{s}=13 $ TeV}},
  \href{http://dx.doi.org/10.1007/JHEP07(2017)121}{\emph{JHEP} {\bf 07} (2017)
  121}, [\href{http://arxiv.org/abs/arXiv:1703.03995}{{\tt arXiv:1703.03995}}].

\bibitem{Aaboud:2017opj}
{ATLAS Collaboration}, \emph{{Search for B-L R-parity-violating top squarks in
  $\sqrt{s}=13$ TeV pp collisions with the ATLAS experiment}},
  \href{http://dx.doi.org/10.1103/PhysRevD.97.032003}{\emph{Phys. Rev. D} {\bf
  97} (2018) 032003}, [\href{http://arxiv.org/abs/arXiv:1710.05544}{{\tt
  arXiv:1710.05544}}].

\bibitem{Sirunyan:2017kqq}
{CMS Collaboration}, \emph{{Search for new phenomena with the $M_{\mathrm
  {T2}}$ variable in the all-hadronic final state produced in proton-proton
  collisions at $\sqrt{s}=13$ TeV}},
  \href{http://dx.doi.org/10.1140/epjc/s10052-017-5267-x}{\emph{Eur. Phys. J.
  C} {\bf 77} (2017) 710}, [\href{http://arxiv.org/abs/arXiv:1705.04650}{{\tt
  arXiv:1705.04650}}].

\bibitem{Sirunyan:2018nkj}
{CMS Collaboration}, \emph{{Search for third-generation scalar leptoquarks
  decaying to a top quark and a $\tau$ lepton at $\sqrt{s}=$ 13 TeV}},
  \href{http://dx.doi.org/10.1140/epjc/s10052-018-6143-z}{\emph{Eur. Phys. J.
  C} {\bf 78} (2018) 707}, [\href{http://arxiv.org/abs/1803.02864}{{\tt
  1803.02864}}].

\bibitem{CMS-PAS-B2G-16-027}
{CMS Collaboration}, \emph{{Search for leptoquarks coupling to third generation
  quarks}}, {\emph{{CMS-PAS-B2G-16-027}} (2018) }.
  \url{https://cds.cern.ch/record/2621420}.

\bibitem{Sirunyan:2017roi}
{CMS Collaboration}, \emph{{Search for standard model production of four top
  quarks with same-sign and multilepton final states in proton–proton
  collisions at $\sqrt{s} = 13\,\text {TeV} $}},
  \href{http://dx.doi.org/10.1140/epjc/s10052-018-5607-5}{\emph{Eur. Phys. J.
  C} {\bf 78} (2018) 140}, [\href{http://arxiv.org/abs/arXiv:1710.10614}{{\tt
  arXiv:1710.10614}}].

\bibitem{Dorsner:2018ynv}
I.~Doršner and A.~Greljo, \emph{{Leptoquark toolbox for precision collider
  studies}}, \href{http://dx.doi.org/10.1007/JHEP05(2018)126}{\emph{JHEP} {\bf
  05} (2018) 126}, [\href{http://arxiv.org/abs/arXiv:1801.07641}{{\tt
  arXiv:1801.07641}}].

\bibitem{Aaboud:2017wqg}
{ATLAS Collaboration}, \emph{{Search for supersymmetry in events with
  $b$-tagged jets and missing transverse momentum in $pp$ collisions at
  $\sqrt{s}=13$ TeV with the ATLAS detector}},
  \href{http://dx.doi.org/10.1007/JHEP11(2017)195}{\emph{JHEP} {\bf 11} (2017)
  195}, [\href{http://arxiv.org/abs/arXiv:1708.09266}{{\tt arXiv:1708.09266}}].

\bibitem{Alloul:2013bka}
A.~Alloul, N.~D. Christensen, C.~Degrande, C.~Duhr and B.~Fuks,
  \emph{{FeynRules 2.0 - A complete toolbox for tree-level phenomenology}},
  \href{http://dx.doi.org/10.1016/j.cpc.2014.04.012}{\emph{Comput. Phys.
  Commun.} {\bf 185} (2014) 2250},
  [\href{http://arxiv.org/abs/arXiv:1310.1921}{{\tt arXiv:1310.1921}}].

\bibitem{Alwall:2014hca}
J.~Alwall, R.~Frederix, S.~Frixione, V.~Hirschi, F.~Maltoni, O.~Mattelaer
  et~al., \emph{{The automated computation of tree-level and next-to-leading
  order differential cross sections, and their matching to parton shower
  simulations}}, \href{http://dx.doi.org/10.1007/JHEP07(2014)079}{\emph{JHEP}
  {\bf 07} (2014) 079}, [\href{http://arxiv.org/abs/arXiv:1405.0301}{{\tt
  arXiv:1405.0301}}].

\bibitem{Ball:2012cx}
R.~D. Ball et~al., \emph{{Parton distributions with LHC data}},
  \href{http://dx.doi.org/10.1016/j.nuclphysb.2012.10.003}{\emph{Nucl. Phys. B}
  {\bf 867} (2013) 244}, [\href{http://arxiv.org/abs/arXiv:1207.1303}{{\tt
  arXiv:1207.1303}}].

\bibitem{Sjostrand:2014zea}
T.~Sjöstrand, S.~Ask, J.~R. Christiansen, R.~Corke, N.~Desai, P.~Ilten et~al.,
  \emph{{An Introduction to PYTHIA 8.2}},
  \href{http://dx.doi.org/10.1016/j.cpc.2015.01.024}{\emph{Comput. Phys.
  Commun.} {\bf 191} (2015) 159},
  [\href{http://arxiv.org/abs/arXiv:1410.3012}{{\tt arXiv:1410.3012}}].

\bibitem{deFavereau:2013fsa}
J.~de~Favereau, C.~Delaere, P.~Demin, A.~Giammanco, V.~Lemaître, A.~Mertens
  et~al., \emph{{DELPHES 3, A modular framework for fast simulation of a
  generic collider experiment}},
  \href{http://dx.doi.org/10.1007/JHEP02(2014)057}{\emph{JHEP} {\bf 02} (2014)
  057}, [\href{http://arxiv.org/abs/arXiv:1307.6346}{{\tt arXiv:1307.6346}}].

\bibitem{Borschensky:2014cia}
C.~Borschensky, M.~Krämer, A.~Kulesza, M.~Mangano, S.~Padhi, T.~Plehn et~al.,
  \emph{{Squark and gluino production cross sections in pp collisions at
  $\sqrt{s}$ = 13, 14, 33 and 100 TeV}},
  \href{http://dx.doi.org/10.1140/epjc/s10052-014-3174-y}{\emph{Eur. Phys. J.
  C} {\bf 74} (2014) 3174}, [\href{http://arxiv.org/abs/arXiv:1407.5066}{{\tt
  arXiv:1407.5066}}].

\bibitem{Read:2002hq}
A.~L. Read, \emph{{Presentation of search results: The CL(s) technique}},
  \href{http://dx.doi.org/10.1088/0954-3899/28/10/313}{\emph{J. Phys. G} {\bf
  28} (2002) 2693}.

\bibitem{Baak:2014wma}
M.~Baak, G.~J. Besjes, D.~C{\^o}t{\'e}, A.~Koutsman, J.~Lorenz and D.~Short,
  \emph{{HistFitter software framework for statistical data analysis}},
  \href{http://dx.doi.org/10.1140/epjc/s10052-015-3327-7}{\emph{Eur. Phys. J.
  C} {\bf 75} (2015) 153}, [\href{http://arxiv.org/abs/arXiv:1410.1280}{{\tt
  arXiv:1410.1280}}].

\bibitem{RooFit:manual}
W.~Verkerke and D.~Kirkb, \emph{Roofit users manual v2.07}, .
  \url{https://roofit.sourceforge.net/docs/RooFit\_Users\_Manual\_2.07-29.pdf}.

\bibitem{James:1994vla}
F.~James, \emph{{MINUIT Function Minimization and Error Analysis: Reference
  Manual Version 94.1}}, .

\bibitem{Ruhr:2016xsg}
{ATLAS Collaboration}, \emph{{Prospects for BSM searches at the high-luminosity
  LHC with the ATLAS detector}},
  \href{http://dx.doi.org/10.1016/j.nuclphysbps.2015.09.094}{\emph{Nucl. Part.
  Phys. Proc.} {\bf 273} (2016) 625}.

\end{thebibliography}\endgroup
\end{document}